\documentclass[fleqn,usenatbib]{mnras}

\usepackage{newtxtext,newtxmath}

\usepackage{amssymb}

\usepackage[T1]{fontenc}

\DeclareRobustCommand{\VAN}[3]{#2}
\let\VANthebibliography\thebibliography
\def\thebibliography{\DeclareRobustCommand{\VAN}[3]{##3}\VANthebibliography}

\usepackage{graphicx}	
\usepackage{amsmath}	
\usepackage{amssymb}	

\usepackage{hyperref}
\hypersetup{
    colorlinks=true,
    linkcolor=cyan,
    citecolor=cyan,
    urlcolor=cyan,
    }

\usepackage{orcidlink}

\usepackage{placeins}

\title[New Galactic symbiotic stars with SALT]{Identification of new Galactic symbiotic stars with SALT - II. New discoveries and characterization of the sample\thanks{Based on observations made with the Southern African Large Telescope (SALT).}}

\author[J. Merc et al.]{
J. Merc,$^{1,2}$\thanks{E-mail: jaroslav.merc@mff.cuni.cz}\orcidlink{0000-0001-6355-2468}
J.~Miko\l{}ajewska,$^{3}$\orcidlink{0000-0003-3457-0020}
K.~I\l{}kiewicz,$^{3}$\orcidlink{0000-0002-4005-5095}
B.~Monard,$^{4}$\orcidlink{0009-0004-6582-0787}
A.~Udalski$^{5}$\orcidlink{0000-0001-5207-5619}
\\
$^{1}$Astronomical Institute, Faculty of Mathematics and Physics, Charles University, V Hole\v{s}ovi\v{c}k{\'a}ch 2, 180 00 Prague, Czech Republic\\
$^{2}$Instituto de Astrof\'isica de Canarias, Calle Vía Láctea, s/n, E-38205 La Laguna, Tenerife, Spain\\
$^{3}$Nicolaus Copernicus Astronomical Center, Polish Academy of Sciences, Bartycka 18, 00–716 Warsaw, Poland\\
$^{4}$Kleinkaroo Observatory, Sint Helena 1B, PO Box 281, Calitzdorp 6660, South Africa\\
$^{5}$Astronomical Observatory, University of Warsaw, Al. Ujazdowskie 4, 00-478 Warszawa, Poland
}

\date{Accepted 2025 December 01. Received 2025 December 01; in original form 2025 October 16}

\pubyear{2025}

\begin{document}
\label{firstpage}
\pagerange{\pageref{firstpage}--\pageref{lastpage}}
\maketitle

\begin{abstract}
We present the continuation of a systematic search for new southern Galactic symbiotic stars, selecting candidates from the SuperCOSMOS H$\alpha$ Survey and 2MASS. Follow-up spectroscopy with the Southern African Large Telescope (SALT) was used to confirm their symbiotic nature and to characterize the cool and hot components of the full sample, including systems from earlier work. We report 14 newly confirmed bona fide symbiotic stars and identify 6 additional strong candidates. Photometric variability was examined using our data and archival light curves from multiple all-sky surveys. Most systems are variable, with the majority showing periodic modulation consistent with orbital motion or pulsations. Possible photometric orbital periods are reported for 19 confirmed and 3 candidate systems, pending spectroscopic confirmation. Eight objects exhibit signs of outburst activity. In one of the systems, multiple brightenings occur at similar orbital phases, closely resembling the evolution of FN Sgr, a symbiotic binary with a magnetic white dwarf. The peculiar variability of another symbiotic star is best explained by dust-obscuration events. These results expand the census of Galactic symbiotic stars.
\end{abstract}



\defcitealias{2014MNRAS.440.1410M}{Paper~I}

\section{Introduction}
Symbiotic stars, binaries consisting of a red giant and a hot companion (white dwarf or, rarely, a neutron star), are among the interacting systems with the longest orbital periods. They are ideal for studying phenomena occurring in a variety of astrophysical sources, that are not yet fully understood and well incorporated in the models of stellar evolution, such as mass transfer and accretion, thermonuclear outbursts, winds, or jets (see the reviews on symbiotic stars by \citealp{2012BaltA..21....5M}, \citealp{2019arXiv190901389M}, and \citealp{2025Galax..13...49M}). Despite the targeted search for the Galactic symbiotic stars is one of the active directions of the current research \citep{2024NatAs...8.1504M} as evidenced by numerous studies in recent years \citep[see, e.g., ][]{2008A&A...480..409C,2010A&A...509A..41C,2013MNRAS.432.3186M,2014MNRAS.440.1410M,2014A&A...567A..49R,2021MNRAS.505.6121M,2021MNRAS.502.2513A,2023MNRAS.519.6044A,2024ApJ...962..126X,2024arXiv241200855L,2025ApJ...987..147C,2025arXiv250720206Z,2025arXiv250620505B,2025MNRAS.tmp.1318B,Merc+Gaia}, their currently known number \citep[<300;][]{2019RNAAS...3...28M,2019AN....340..598M,Merc+NODSV2025} is still significantly lower than any estimate of the population size presented in the literature \citep[ranging from a few thousands to a few hundred thousands;][]{1984PASA....5..369A,1993ApJ...407L..81K,2003ASPC..303..539M,2025A&A...698A.155L}. 

Given that a large fraction of symbiotic stars show very strong emission lines, one of the methods to select promising candidates is to employ large photometric surveys conducted in the H$\alpha$ filter. Several new discoveries were achieved thanks to the INT Photometric H$\alpha$ survey \citep[IPHAS; ][]{2005MNRAS.362..753D}, see \citet{2008A&A...480..409C,2010A&A...509A..41C} and \citet{2014A&A...567A..49R}, and the AAO/UKST SuperCOSMOS H$\alpha$ Survey \citep[SHS;][]{2005MNRAS.362..689P}. In case of a later survey, \citet[][]{2013MNRAS.432.3186M} studied a $\sim$35 deg$^2$ region towards the Galactic bulge, discovering 20 new symbiotic stars and another 15 candidates. \citet[][hereafter \citetalias{2014MNRAS.440.1410M}]{2014MNRAS.440.1410M} used the SHS survey to 
 search for new Galactic southern symbiotic stars. The spectroscopic follow-up using the Southern African Large Telescope (SALT) resulted in the discovery of 12 new and 3 possible symbiotic stars.

In this paper, we continue in the systematic survey presented in \citetalias{2014MNRAS.440.1410M} and present the follow-up of additional sources selected using the criteria discussed in Sect. 2 of \citetalias{2014MNRAS.440.1410M} based on the SHS H$\alpha$ data and 2MASS near-infrared observations \citep{2006AJ....131.1163S}. In addition to presenting new discoveries, we analyze some of the basic stellar parameters of the components of symbiotic stars from our sample (new and the ones from \citetalias{2014MNRAS.440.1410M}), study the variability of the stars, and, if possible, discuss their orbital parameters and pulsational properties of their cool components.

The paper is organized as follows: Section~\ref{sec:obs} describes the spectroscopic data and archival photometry used in this work. Section~\ref{sec:new} presents the discovery of 14 new symbiotic stars and 6 strong candidates. In Section~\ref{sec:parameters}, we analyze the parameters of the symbiotic components in the full SALT sample, while Section~\ref{sec:variability} discusses their photometric variability with the aim of identifying orbital and pulsational periods. Section~\ref{sec:individual} focuses on individual noteworthy objects, and Section~\ref{sec:conclusions} summarizes our findings.

\section{Observational data}\label{sec:obs}
\subsection{Optical spectroscopy}
The new symbiotic candidates were observed spectroscopically using the Robert Stobie Spectrograph \citep[RSS;][]{2003SPIE.4841.1463B,2003SPIE.4841.1634K} on the SALT telescope \citep{2006SPIE.6267E..0ZB,2006MNRAS.372..151O} under programmes 2013-2-RSA\_POL-001 (PI: Miszalski), 2014-1-RSA\_POL-001 (PI: Miszalski), 2015-1-SCI-011 (PI: Miszalski), and 2021-2-SCI-009 
(PI: Miko\l{}ajewska). The SALT/RSS spectra for symbiotic stars whose discovery was presented in \citetalias{2014MNRAS.440.1410M} were obtained under programme 2013-1-RSA\_POL-001 (PI: Miszalski). The details on the RSS configuration and processing of the spectra can be found in Section 3 of \citetalias{2014MNRAS.440.1410M}. The log of new spectroscopic observations is listed in Tab. \ref{tab:salt_spectra} and shows the 2MASS identification of the object, observing date, and exposure time of the longer of two exposures of the same star that was always taken after a short 30 or 60s exposure to measure H$\alpha$ unsaturated. The same information for symbiotic stars from \citetalias{2014MNRAS.440.1410M} that are also analyzed in this work is not repeated here, as it is presented in table 1 of the aforementioned paper. 

\begin{table}
	\centering
	\caption{Log of SALT/RSS observations. Only the exposure times of longer exposures are given (see text).}
	\label{tab:salt_spectra}
\begin{tabular}{lcr}
\hline
Name (2MASS~J) & Date (dd-mm-yy) & Exposure (s) \\
\hline
06503882-0006394 & 09-11-2013 & 1200 \\
10532463-6044518 & 19-12-2013 & 1200 \\
12283699-6519204 & 15-01-2014 & 1200 \\
14031865-5809349 & 08-04-2022 & 462 \\
16284838-4010161 & 01-05-2014 & 1170 \\
16472941-3612388 & 02-05-2014 & 1170 \\
16592100-4517173 & 14-05-2014 & 1170 \\
17160302-3322285 & 01-05-2014 & 1170 \\
17210951-4416404 & 13-06-2014 & 1800 \\
17303001-3049372 & 13-07-2014 & 1170 \\
17341102-3850446 & 14-07-2014 & 1170 \\
17370406-3324539 & 14-05-2014 & 1170 \\
17370603-2500098 & 22-06-2014 & 1170 \\
17371139-3350500 & 21-05-2014 & 1170 \\
17374702-2501120 & 11-08-2014 & 1170 \\
17435611-2506254 & 12-06-2015 & 1200 \\
17505978-3012473 & 29-06-2014 & 1170 \\
17562573-1631486 & 16-06-2014 & 790 \\
18002542-1126324 & 11-05-2014 & 1170 \\
18075073-2516427 & 24-05-2015 & 1200 \\
18155753-0837357 & 11-05-2014 & 1170 \\
\hline
\end{tabular}
\end{table}

\subsection{Photometry}

To study the variability of the target stars, we have searched for available data from the All-Sky Automated Survey for Supernovae \citep[ASAS-SN;][]{2014ApJ...788...48S, 2017PASP..129j4502K}, the Optical Gravitational Lensing Experiment survey \citep[OGLE~III and IV; ][]{2008AcA....58...69U,2015AcA....65....1U}, the Zwicky Transient Facility survey \citep[ZTF;][]{2019PASP..131a8003M}, and the Asteroid Terrestrial-impact Last Alert System (ATLAS) project \citep{2018PASP..130f4505T,2020PASP..132h5002S} obtained from the ATLAS Forced Photometry server \citep{2021TNSAN...7....1S}. We have also searched for detections of the sources on the digitized photographic plates from the Harvard College Observatory from the DASCH (Digital Access to a Sky Century at Harvard) archive \citep{2010AJ....140.1062L}. For some objects from \citetalias{2014MNRAS.440.1410M}, we have also collected our photometry in the $V$ and $I$ filters using the 35 cm Schmidt–Cassegrain and Ritchey-Chretien telescopes equipped with SBIG CCD cameras at the Kleinkaroo Observatory in South Africa.

To search for periodicities in the light curves, we performed period analysis on individual datasets using the Lomb-Scargle method \citep{1976Ap&SS..39..447L,1982ApJ...263..835S}, as well as the multi-band Lomb-Scargle variant \citep{2015ApJ...812...18V}, both implemented in the {\tt astropy} Python package \citep{2013A&A...558A..33A,2018AJ....156..123A,2022ApJ...935..167A}.

\section{New symbiotic stars}\label{sec:new}

\begin{table*}
	\centering
	\caption{The list of the new and possible symbiotic stars detected in this work. Note that if the SIMBAD identifier is the same as the 2MASS name of the star, we do not repeat it in whole. The parallax ($\varpi$) is from \textit{Gaia} DR3 \citep{2023A&A...674A...1G}. The distances listed in the table are photogeometric distances from \citet{2021AJ....161..147B}.}
	\label{tab:basic_data}
\begin{tabular}{lllrrrrr}
\hline

Name (2MASS~J) & SIMBAD name & \textit{Gaia} DR3 & OGLE IV & $\ell$  & $b$  & $\varpi$  & d  \\
 &  &  &  & (\degr) & (\degr) & (mas) & (kpc) \\
\hline
\textit{Bona-fide} &  &  &  &  &  &  &  \\
16284838-4010161 & 2MASS~J162.. & 6017140952831727616 & GD2159.10.14478 & 341.5504 & 5.8712 & 0.07$\pm$0.03 & 6.6 \\
16472941-3612388 & 2MASS~J164.. & 6019738480355216512 & - & 346.9756 & 5.7579 & -0.01$\pm$0.02 & 9.1 \\
17160302-3322285 & Terz V 4021 & 5979017105159069312 & BLG918.21.90 & 352.8160 & 2.8452 & -0.02$\pm$0.05 & 7.4 \\
17303001-3049372 & IRAS 17272-3047 & 4058383390530172288 & BLG662.03.62513 & 356.6487 & 1.7567 & 0.56$\pm$0.26 & 4.4 \\
17341102-3850446 & 2MASS~J173.. & 5961813076738022144 & BLG674.19.27171 & 350.3400 & -3.2534 & 0.02$\pm$0.08 & 6.1 \\
17370603-2500098 & 2MASS~J173.. & 4062231784311861376 & - & 2.3399 & 3.6885 & 0.00$\pm$0.07 & 6.8 \\
17371139-3350500 & 2MASS~J173.. & 4053683351942447104 & - & 354.8806 & -1.0702 & 0.12$\pm$0.14 & 6.2 \\
17374702-2501120 & 2MASS~J173.. & 4062219066862909952 & BLG714.25.14317 & 2.4080 & 3.5484 & 0.12$\pm$0.06 & 6.3 \\
17435611-2506254 & 2MASS~J174.. & 4067943811590065536 & BLG633.07.22889 & 3.0705 & 2.3194 & 0.32$\pm$0.07 & 4.2 \\
17505978-3012473 & MSX6C G359.5102-01.6611 & 4056361899741275392 & BLG501.05.167800 & 359.5094 & -1.6603 & 0.16$\pm$0.08 & 6.8 \\
17562573-1631486 & IRAS 17535-1631 & 4144709995465485696 & BLG837.03.36067 & 11.9486 & 4.1915 & 0.19$\pm$0.04 & 3.9 \\
18002542-1126324 & ZTF J180025.42-112632.5 & 4151813325933260672 & BLG840.30.24 & 16.8707 & 5.8606 & -0.01$\pm$0.04 & 8.3 \\
18075073-2516427 & OGLE BLG-LPV-214983 & 4065672156232576640 & BLG580.22.74954 & 5.6472 & -2.4394 & -0.02$\pm$0.07 & 6.4 \\
18155753-0837357 & 2MASS~J181.. & 4157938533401345920 & - & 21.1922 & 3.8577 & 0.05$\pm$0.14 & 6.4 \\
 &  &  &  &  &  &  &  \\
\textit{Possible} &  &  &  &  &  &  &  \\
06503882-0006394 & IRAS 06480-0003 & 3113251991344790144 & GD1730.11.5882 & 212.9406 & -0.2265 & 0.25$\pm$0.08 & 3.1 \\
10532463-6044518 & WRAY 15-678 & 5338200425455290368 & GD1369.18.25889 & 289.0028 & -1.1085 & 0.00$\pm$0.06 & 6.3 \\
12283699-6519204 & WRAY 16-113 & 5860998619435294080 & GD1304.13.26309 & 300.5507 & -2.5562 & 0.48$\pm$0.04 & 1.9 \\
16592100-4517173 & 2MASS~J165.. & 5964068591723229184 & GD1082.31.90 & 341.3460 & -1.7121 & -0.04$\pm$0.07 & 5.7 \\
17210951-4416404 & 2MASS~J172.. & 5953276610229325056 & BLG987.08.13 & 344.4560 & -4.2310 & 0.10$\pm$0.03 & 5.6 \\
17370406-3324539 & 2MASS~J173.. & 4053748429286984320 & BLG661.25.39044 & 355.2317 & -0.8168 & 0.04$\pm$0.14 & 6.9 \\
\hline
\end{tabular}
\end{table*}

\begin{figure*}
\centering
 \includegraphics[width=0.8\textwidth]{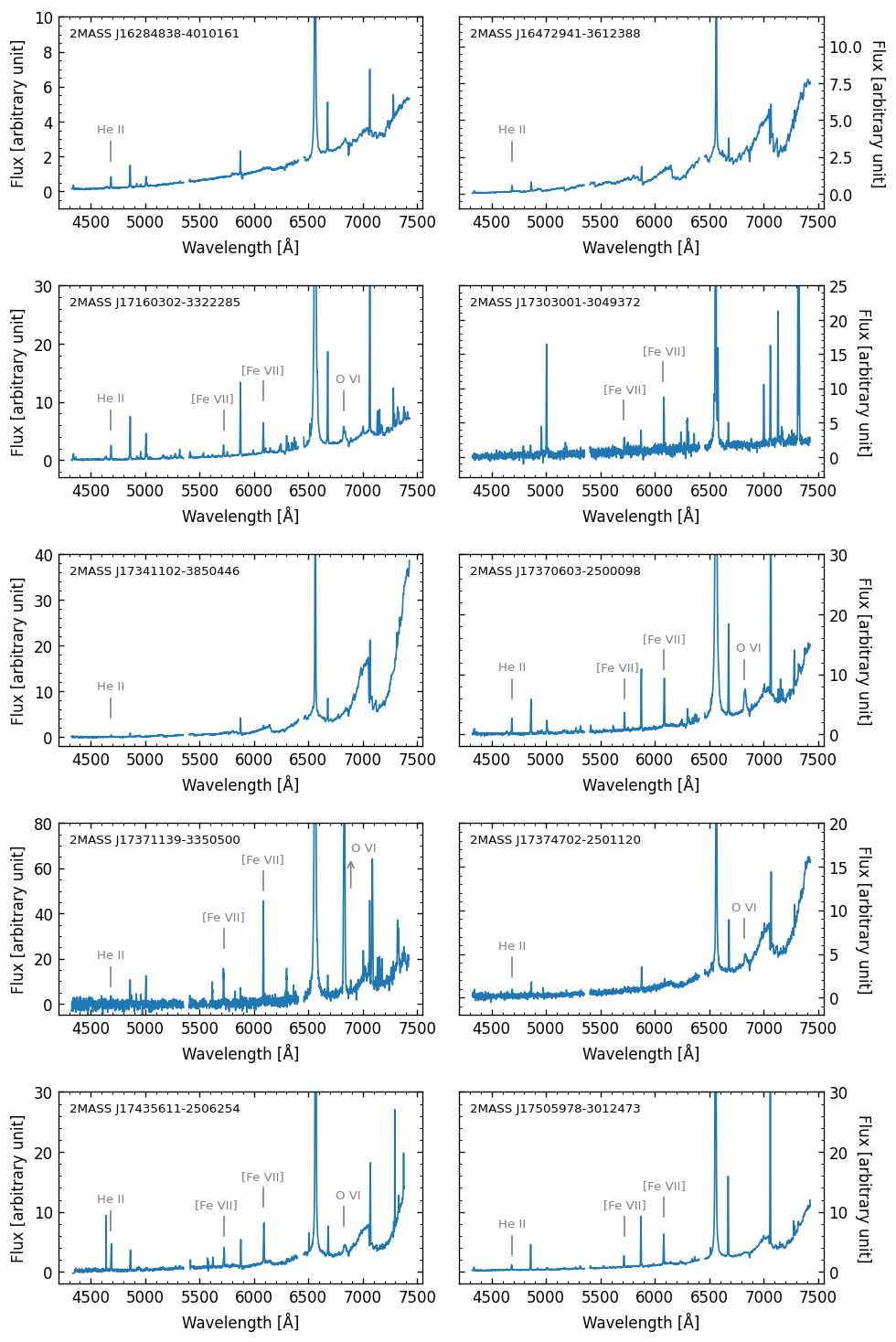}
 \caption{SALT/RSS spectra of new symbiotic stars. The spectra are not dereddened, and due to limitations in absolute flux calibration due to the moving pupil design of SALT, they are normalized by the average continuum value measured in the region 6\,200-6\,300 \AA{} \citep[see][]{2014MNRAS.440.1410M}. The identification of selected emission lines is shown in gray. Most of the unmarked emission lines are of \ion{H}{i} and \ion{He}{i}.}
 \label{fig:spectra_1}
 \end{figure*}

 \begin{figure*}
  
 \centering
 \includegraphics[width=0.8\textwidth]{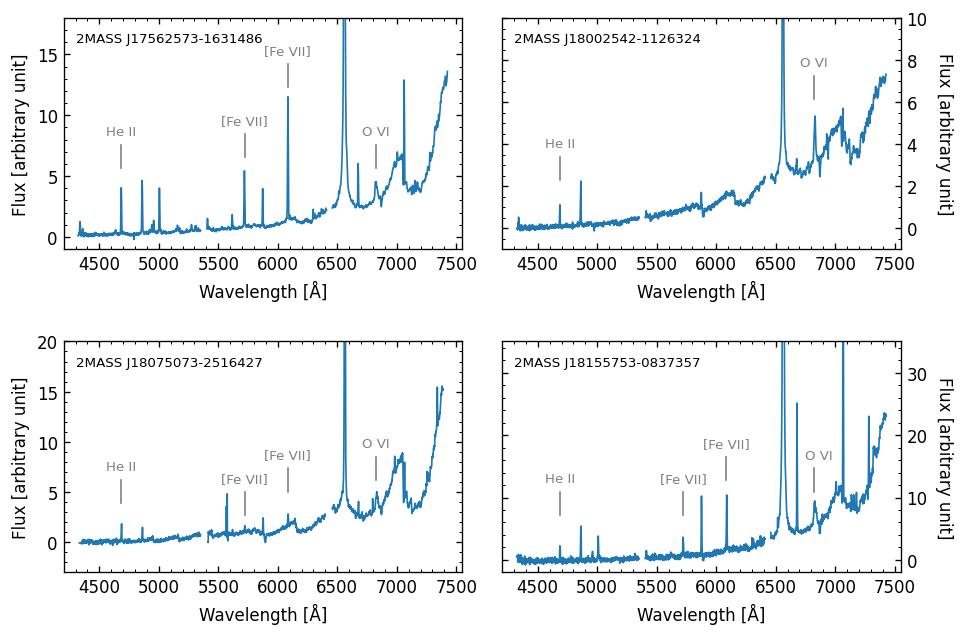}

 \contcaption{SALT/RSS spectra of new symbiotic stars.}

 \label{fig:spectra_2}
 \end{figure*}

\begin{table*}
 \setlength{\tabcolsep}{5.1pt}
	\centering
	\caption{\textit{Gaia} DR3 and 2MASS near-IR magnitudes of new and possible symbiotic stars detected in this work, together with the adopted value of interstellar extinction (see the text for details), infrared symbiotic type, spectral type of the giant, and the effective temperature obtained from the calibration of \citet[][]{1999AJ....117..521V}.}
	\label{tab:2mass}
\begin{tabular}{lrrr|rrrrccr}
\hline

Name & $G_{\rm BP}$ & $G$ & $G_{\rm RP}$ & $J$ & $H$ & $K_s$ & E$_{\rm(B-V)}$ & IR & Giant & T$_{\rm eff}$ \\
 & (mag) & (mag) & (mag) & (mag) & (mag) & (mag) & (mag) & type & SpT & (K) \\ \hline
\textit{Bona-fide} &  &  &  &  &  &  &  &  &  &  \\
16284.. & 16.7 & 15.2 & 13.9 & 11.38 & 10.27 & 9.72 & 1.1 & S & M1 & 3804 \\
16472.. & 15.8 & 14.0 & 12.7 & 10.46 & 9.33 & 8.98 & 1.0 & S & M3 & 3586 \\
17160.. & 16.8 & 14.9 & 13.5 & 9.50 & 8.04 & 7.34 & 1.6 & S & M0 & 3914 \\
17303.. & 18.7 & 17.4 & 15.9 & 12.32 & 10.29 & 8.47 & 1.6 & D & - & - \\
17341.. & 18.3 & 15.0 & 13.4 & 10.10 & 8.78 & 8.27 & 1.5 & S & M5 & 3367 \\
17370.. & 18.1 & 15.5 & 14.0 & 10.11 & 8.65 & 8.00 & 1.4 & S & M2 & 3695 \\
17371.. & 19.1 & 16.2 & 14.5 & 10.20 & 8.87 & 8.09 & 1.8 & S & M0 & 3914 \\
17374.. & 18.5 & 15.7 & 14.1 & 10.65 & 9.30 & 8.72 & 1.6 & S & M3 & 3586 \\
17435.. & 19.3 & 15.9 & 14.3 & 10.44 & 9.07 & 8.37 & 1.4 & S & M3 & 3586 \\
17505.. & 17.8 & 15.1 & 13.5 & 9.73 & 8.43 & 7.80 & 1.2 & S & M2.5 & 3641 \\
17562.. & 15.2 & 13.2 & 11.8 & 8.72 & 7.53 & 6.95 & 0.9 & S & M4 & 3476 \\
18002.. & 17.0 & 15.1 & 13.9 & 11.31 & 10.23 & 9.83 & 0.9 & S & M2 & 3695 \\
18075.. & 18.7 & 15.8 & 14.3 & 11.30 & 10.02 & 9.52 & 1.1 & S & M3 & 3586 \\
18155.. & 17.9 & 15.4 & 13.8 & 9.99 & 8.56 & 7.81 & 1.4 & S & M2 & 3695 \\
 &  &  &  &  &  &  &  &  &  &  \\
\textit{Possible} &  &  &  &  &  &  &  &  &  &  \\
06503.. & 17.5 & 16.8 & 16.0 & 13.60 & 11.95 & 10.56 & 0.5 & D? & - & - \\
10532.. & 17.3 & 16.8 & 15.8 & 13.38 & 11.17 & 9.26 & 2.0 & D? & - & - \\
12283.. & 14.8 & 14.6 & 13.5 & 12.18 & 10.90 & 9.65 & 0.3 & D? & - & - \\
16592.. & 18.6 & 15.5 & 14.0 & 10.37 & 8.96 & 8.40 & 1.5 & S? & M2 & 3695 \\
17210.. & 16.8 & 14.5 & 13.1 & 10.39 & 9.16 & 8.71 & 1.1 & S? & M4 & 3476 \\
17370.. & 20.2 & 17.1 & 15.5 & 11.13 & 9.62 & 8.87 & 2.1 & S? & - & - \\
\hline
\end{tabular}
\end{table*}

 \begin{figure*}
\centering
 \includegraphics[width=0.8\textwidth]{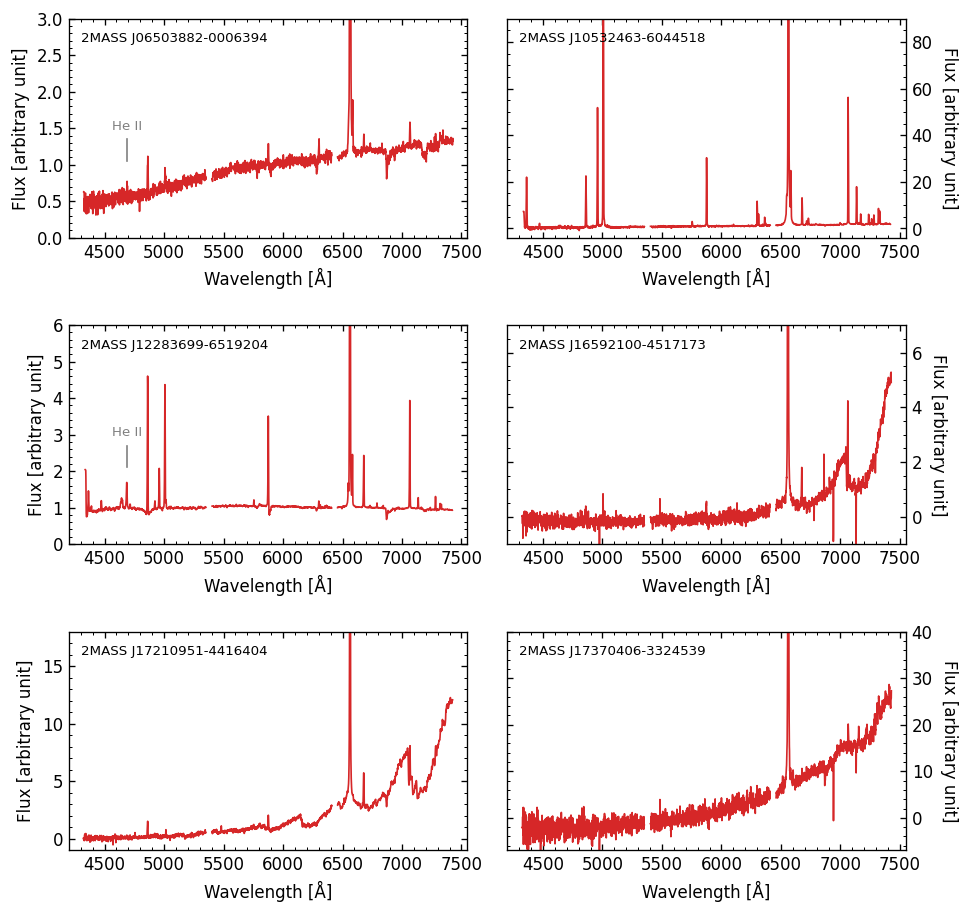}
 \caption{SALT/RSS spectra of possible symbiotic stars.}
 \label{fig:spectra_possible}
 \end{figure*}

In this work, we present the discovery of 14 new symbiotic stars, and we also discuss 6 strong symbiotic candidates (see Table \ref{tab:basic_data}). To classify an object as a symbiotic star, we adopted the criteria of \citet{2000A&AS..146..407B} and \citet{2013MNRAS.432.3186M}, namely we required the simultaneous presence of the late-type giant features in the spectrum (e.g., TiO, \ion{Ca}{ii} and other absorption lines) and detection of strong emission lines of H I and He I, together with emission lines with ionization potential of at least 35 eV (e.g., [\ion{O}{iii}], \ion{He}{ii}). The detection of Raman-scattered \ion{O}{vi} lines at 6\,825 and 7\,082~\AA{} \citep{1989A&A...211L..31S} is a sufficient but not necessary condition for the symbiotic classification. These lines are observed in about half of the confirmed symbiotic systems \citep[e.g., ][]{2019ApJS..240...21A,2019RNAAS...3...28M}.

The SALT spectra of new symbiotic stars are shown in Fig.~\ref{fig:spectra_1}. All of the newly discovered symbiotic stars show strong emission lines of \ion{H}{i} and \ion{He}{i}, and also of \ion{He}{ii}, [\ion{Fe}{vii}], and nine of them show also Raman-scattered \ion{O}{vi} lines. They are very intense in 2MASS~J17371139-3350500. Two of the new bona-fide symbiotic stars were previously classified as H$\alpha$ emission sources by \citet{2003AN....324..437K}, namely 2MASS~J17370603-2500098 ([KW2003] 40) and 2MASS~J17505978-3012473 ([KW2003] 69).

SALT spectra of the candidates are shown in Fig.~\ref{fig:spectra_possible}. The optical spectra do not fully fulfill the criteria for definitive symbiotic classification (either due to the lack of highly ionized emission lines in the spectra or not definitive evidence for the red giant in the optical spectra). Several of the current symbiotic candidates were considered as possible PNe in the literature previously: 2MASS~J06503882-0006394 was classified as a possible PN in \citet{2020A&A...638A..21V}, 2MASS~J10532463-6044518 was classified as an H$\alpha$ emission line star by \citet{1966PhDT.........3W} and \citet{1990AJ....100..793S}, and as possible PN by \citet{2020A&A...638A..21V}. 2MASS~J12283699-6519204 was previously classified as a likely PN by \citet{1966PhDT.........3W}. However, the red 2MASS colors of these three candidates with PN-like spectra (marked D? in Table \ref{tab:2mass}) are typical for D-type symbiotic stars and not for PNe. Furthermore, the approximate positions of the two D-type candidates, 2MASS~J10532463-6044518 and ~J12283699-6519204, in the [\ion{O}{iii}] diagnostic diagram\footnote{The flux in the H$\gamma$ line could not be directly measured from the SALT spectra, as it lies near the blue edge of the spectral range; however, an approximate flux was derived from the red wing of the line.} \citep{1995PASP..107..462G,2017A&A...606A.110I} support a symbiotic rather than PN classification. No H$\gamma$ or [\ion{O}{iii}] $\lambda$4\,363 emission is detectable in 2MASS~J06503882-0006394, and therefore this diagnostic cannot be applied to that object.

\section{Stellar parameters of the SALT sample}\label{sec:parameters}
All but one of the new symbiotic stars are classified as infrared type S ("stellar"; see Table \ref{tab:2mass}, which also lists 2MASS magnitudes of target stars), with the red giant's presence clearly visible in our optical spectra (Fig.~\ref{fig:spectra_1}). 2MASS~J17303001-3049372 exhibits the typical optical spectrum of a D-type symbiotic star, showing, for example, much stronger [\ion{O}{iii}]/\ion{H}{i} ratios and prominent [\ion{N}{ii}] emission, features not typically seen in S-types \citep[see, e.g.,][]{2017A&A...606A.110I}. This classification is further supported by the infrared excess in its SED, characteristic of D-type symbiotics.

To estimate the spectral types of the cool companions in new S-type symbiotic stars and candidates, we adopted the same methodology as in \citetalias{2014MNRAS.440.1410M} and \citet{2013MNRAS.432.3186M}, which utilizes the TiO indices from \citet{1987AJ.....93..938K}; see their equations 1 and~2. The resulting spectral types are provided in Table \ref{tab:2mass}, and the corresponding effective temperatures were derived using the calibration by \citet{1999AJ....117..521V}. Similar information for symbiotic stars from \citetalias{2014MNRAS.440.1410M} is available in Table \ref{tab:2mass_paperI}. It should be noted that the optical spectral types represent upper limits, as the depths of the TiO bands are likely diluted by the blue (predominantly nebular) continuum. Near-infrared spectra of symbiotic stars typically indicate later spectral types, and for the same reason, the effective temperatures derived from the optical spectra should also be regarded as upper limits.

We have not estimated the luminosities of these symbiotic giants, as the distances to these objects remain uncertain. Table \ref{tab:basic_data} lists photogeometric distances from \citet{2021AJ....161..147B}, based on \textit{Gaia} EDR3/DR3 data. However, for these distant symbiotic systems, the parallaxes are unreliable, with many being negative, despite RUWE values or goodness-of-fit statistics from \textit{Gaia} appearing reasonable in some cases.

The hot components are typically not directly visible in the optical spectra of symbiotic systems but are evidenced by highly ionized emission lines. These emission lines, influenced by the radiation of the hot component, provide indirect means to estimate its parameters.

 \begin{table*}
	\centering
     \setlength{\tabcolsep}{5.1pt}
	\caption{Fluxes of emission lines in the spectra of new and possible symbiotic stars relative to the flux of H$\beta$ line, together with the ion with the maximum ionization potential and the estimates of the hot component temperature.}
	\label{tab:nebular}
\begin{tabular}{llllllllllllllll}
\hline

Name & H$\gamma$ & {[}\ion{O}{iii}{]}  & \ion{He}{ii}  & H$\beta$ & {[}\ion{O}{iii}{]}  & {[}\ion{Fe}{vii}{]} & \ion{He}{i}  & {[}\ion{Fe}{vii}{]} & H$\alpha$ & \ion{He}{i}   & \ion{O}{vi}  & \ion{He}{i}  & Ion & T$^{a}_{\rm h}$ & T$^{b}_{\rm h}$ \\
 & {\scriptsize 4340 \AA} & {\scriptsize 4363 \AA}& {\scriptsize 4686 \AA} & {\scriptsize 4861 \AA}& {\scriptsize 5007 \AA} & {\scriptsize 5721 \AA} & {\scriptsize 5876 \AA}& {\scriptsize 6086 \AA}& {\scriptsize 6563 \AA}& {\scriptsize 6678 \AA}& {\scriptsize 6825 \AA}& {\scriptsize 7065 \AA}&  & (10$^3$ K) & (10$^3$ K) \\ \hline
\textit{Bona-fide} &  &  &  &  &  &  &  \\
16284.. & 27 & 0 & 55 & 100 & 30 & 0 & 38 & 2 & 1414 & 61 & 20 & 53 & O$^{+5}$ & 114 & 156 \\
16472.. & 30 & 7 & 72 & 100 & 0 & 6 & 71 & 6 & 1235 & 94 & 0 & 116 & Fe$^{+6}$: & 99 & 171 \\
17160.. & 28 & 7 & 38 & 100 & 46 & 10 & 49 & 21 & 1043 & 35 & 19 & 74 & O$^{+5}$ & 114 & 138 \\
17303.. & 23 & 7 & 40 & 100 & 1247 & 79 & 75 & 205 & 4583 & 59 & 0 & 198 & Fe$^{+6}$ & 99 & 141 \\
17341.. & - & 0 & 55 & 100 & 18 & 8 & 126 & 19 & 2101 & 119 & 0 & 225 & Fe$^{+6}$: & 99 & 156 \\
17370.. & 25 & 8 & 48 & 100 & 30 & 22 & 61 & 46 & 2896 & 58 & 40 & 116 & O$^{+5}$ & 114 & 149 \\
17371.. & - & 0 & 23 & 100 & 95 & 47 & 17 & 125 & 5266 & 18 & 453 & 44 & O$^{+5}$ & 114 & 119 \\
17374.. & - & 0 & 74 & 100 & 0 & 9 & 54 & 8 & 1727 & 70 & 47 & 71 & O$^{+5}$ & 114 & 173 \\
17435.. & - & 4 & 155 & 100 & 5 & 47 & 47 & 75 & 1415 & 31 & 30 & 63 & O$^{+5}$ & 114 & 227 \\
17505.. & 24 & 0 & 26 & 100 & 6 & 24 & 79 & 52 & 2571 & 82 & 0 & 143 & Fe$^{+6}$ & 99 & 123 \\
17562.. & 32 & 14 & 96 & 100 & 69 & 60 & 32 & 114 & 2026 & 27 & 36 & 50 & O$^{+5}$ & 114 & 190 \\
18002.. & 19 & 0 & 50 & 100 & 0 & 0 & 16 & 3 & 1692 & 15 & 88 & 19 & O$^{+5}$ & 114 & 151 \\
18075.. & - & 0 & 151 & 100 & 0 & 24 & 58 & 40 & 2031 & 43 & 123 & 73 & O$^{+5}$ & 114 & 225 \\
18155.. & - & 0 & 44 & 100 & 59 & 26 & 64 & 63 & 3526 & 84 & 54 & 148 & O$^{+5}$ & 114 & 145\\
 &  &  &  &  &  &  &  \\
\textit{Possible} &  &  &  &  &  &  &  \\
06503.. & 0 & 0 & 0 & 100 & 57 & 0 & 37 & 0 & 1757 & 26 & 0 & 33 & O$^{+2}$ & 35 & - \\
10532.. & - & 273 & 0 & 100 & 564 & 0 & 31 & 0 & 718 & 6 & 0 & 20 & O$^{+2}$ & 35 & - \\
12283.. & - & 22 & 28 & 100 & 88 & 0 & 51 & 0 & 791 & 28 & 0 & 51 & He$^{+2}$ & 54 & 127 \\
16592.. & - & 0 & 0 & 100 & 150 & 0 & 84 & 0 & 6635 & 184 & 0 & 288 & O$^{+2}$ & 35 & - \\
17210.. & 23 & 0 & 0 & 100 & 0 & 0 & 32 & 0 & 1172 & 73 & 0 & 55 & He$^{+1}$ & 25 & - \\
17370.. & - & 0 & 0 & 100 & 0 & 0 & 2 & 0 & 450 & 2 & 0 & 6 & He$^{+1}$ & 25 & - \\
\hline
\end{tabular}

 \medskip
    \begin{minipage}{\linewidth}
    \textbf{Notes.} $^{a}$Temperature of the hot component estimated from the maximum ionization potential observed in the optical spectrum.
$^{b}$Temperature of the hot component estimated using the method of \citet{1981ASIC...69..517I}. For further details, see the main text.
    \end{minipage}
\end{table*}

The lower limit of the hot component's temperature is determined by the maximum ionization potential observed in the spectrum (see Table \ref{tab:nebular}) and the relation $T$ [10$^3$ K] $\sim$ $IP_{\text{max}}$ [eV] proposed by \citet{1994A&A...282..586M}. For the newly discovered symbiotic stars, the emission lines with maximum ionization potential are [\ion{Fe}{vii}] or Raman-scattered \ion{O}{vi} lines, implying hot component temperatures exceeding 99 kK or 114 kK, respectively. The same holds for symbiotic stars from \citetalias{2014MNRAS.440.1410M} (see Table \ref{tab:nebular_paperI}), except for 2MASS~J17334728-2719266, which only shows \ion{He}{ii}.

The temperature of the ionizing photon source, assuming case B recombination, can also be derived following the method of \citet{1981ASIC...69..517I}, which involves fluxes of \ion{He}{i} $\lambda$4\,471, \ion{He}{ii} $\lambda$4\,868, and H$\beta$ emission lines. These temperatures are provided in Tables \ref{tab:nebular} and \ref{tab:nebular_paperI}. For most of the new symbiotic stars, the \ion{He}{i} $\lambda$4\,471 flux is negligible (or very low and difficult to precisely measure due to the low S/N in the blue region), so it was neglected in the calculations, as done by \citet{2006ApJ...636.1002S} for Z And, and \citet{2016MNRAS.456.2558L} and \citet{2017gacv.workE..60M} for AG Dra.

Tables \ref{tab:nebular} and \ref{tab:nebular_paperI} list the fluxes of \ion{He}{ii} $\lambda$4\,868 and H$\beta$ lines, along with other prominent lines detected in the spectra. These fluxes were dereddened using the reddening values from Tables \ref{tab:basic_data} and \ref{tab:basic_data_paperI}, respectively, assuming the \citet{1989ApJ...345..245C} reddening law and adopting a total-to-selective absorption ratio R = 3.1. The adopted reddening values correspond to the total reddening in the direction of the target stars at their distances, obtained using the {\tt mwdust} code \citep{2016ApJ...818..130B} from combined 3D dust maps of \citet{2003A&A...409..205D}, \citet{2006A&A...453..635M}, and \citet{2019ApJ...887...93G}.

The luminosity of the hot component can also be evaluated from optical emission lines (e.g., equation 8 of \citeauthor{1991AJ....101..637K}, \citeyear{1991AJ....101..637K}, and equations 6 and 7 of \citeauthor{1997A&A...327..191M}, \citeyear{1997A&A...327..191M}). However, since the SALT spectra analyzed in this work are only relative, this approach is not applicable here.  

\section{Photometric variability}\label{sec:variability}

To study the variability of newly confirmed and candidate symbiotic stars identified in this work and in \citetalias{2014MNRAS.440.1410M}, we compiled photometric data from various surveys. A large fraction of the objects have useful light curves available from the ATLAS survey\footnote{Note that the data were obtained in non-standard filters: orange ($\sim$5\,582--8\,249 \AA{}, $\lambda_{\rm eff} = 6\,630$ \AA{}) and cyan ($\sim$4\,157--6\,556 \AA{}, $\lambda_{\rm eff} = 5\,182$ \AA{}).}, and several are also covered by OGLE-IV, although in some cases the light curves are too short to allow reliable period analysis. Only a handful of sources have ASAS-SN data of sufficient quality, largely because most of the SALT sample is relatively faint and/or located in crowded fields. On the other hand, ZTF data are generally quite useful; however, the survey only observes objects with declinations above -30\textdegree{}, which excludes a significant portion of the sample. 

Light curves of bona-fide and candidate symbiotic stars from this work are presented in Fig.~\ref{fig:photometry} and Fig.~\ref{fig:photometry5}, respectively. The same information for objects from \citetalias{2014MNRAS.440.1410M} is shown in Fig.~\ref{fig:photometry8} and Fig.~\ref{fig:photometry10}. With the exception of a few candidate symbiotic stars, all objects exhibit variability on some timescale, which is clearly periodic in most cases. Most of these systems are already classified as long-period variables or candidates in the literature, largely thanks to the \textit{Gaia} DR3 variability data \citep[see][]{2023A&A...674A..15L,2023A&A...674A..14R,2023A&A...674A..13E}. A summary of the main results of our analysis is provided below, while individual noteworthy systems are discussed in Sect. \ref{sec:individual}.

Tables \ref{tab:variability} and \ref{tab:variability_paperI} list the periodicities detected in the objects from this work and from \citetalias{2014MNRAS.440.1410M}, respectively. In many cases, the dominant periodic signal is interpreted as the orbital period of the binary system, based on the characteristic timescales, amplitudes, color variations, and light-curve morphology (see more detailed discussion for individual objects in Sect. \ref{sec:individual}). In total, we identify tentative orbital periods for 19 bona-fide and 3 candidate symbiotic stars, ranging from 384 to 1518 days, a range typical for S-type symbiotic systems \citep[e.g.,][]{2013AcA....63..405G}. We emphasize that these values require confirmation through spectroscopic monitoring. Several light curves also show hints of eclipses.

In addition, possible pulsations of the cool component were detected in some cases, particularly after subtracting the orbital signal and/or in the redder bands. In four objects, the dominant variability appears to arise from Mira pulsations of the cool component. Among them, 2MASS~J17303001-3049372 (this work) and 2MASS~J16422739-4133105 (\citetalias{2014MNRAS.440.1410M}) are the only two confirmed D-type symbiotic stars in our combined sample. The other two, 2MASS~J17371139-3350500 (this work) and 2MASS~J17334728-2719266 (\citetalias{2014MNRAS.440.1410M}), exhibit characteristics consistent with S-type systems. While none of the D-type candidates from either study show clear Mira-like pulsations, most are only covered by OGLE-IV; only 2MASS~J06503882-0006394 has additional data from ATLAS and ZTF, and it shows complex variability.

Finally, eight of the confirmed symbiotic stars display signatures of outburst activity in their light curves. Seven of these were identified from modern survey data, while in one case, the outburst was detected on historical DASCH photographic plates. 

\begin{table}
	\centering
	\caption{Variability of new and possible symbiotic stars.}
	\label{tab:variability}

\begin{tabular}{lrrrcc}
\hline
Name & P$_{\rm orb}$ & P$_{\rm pul}$ & P$_{\rm other}$ & Outbursts & Eclipses \\
 & (days) & (days) & (days) &  &  \\ \hline
\textit{Bona-fide} &  &  &  &   \\
16284.. & 741 &  &  & & \\
16472.. & 502: & 81: & 901: & &  \\
17160.. & 1518: & 142 & & Yes &  \\
17303.. &  & 548 &  & & \\
17341.. & 694: &  &  &  & \\
17370.. & 778 &  &  &  \\
17371.. &  & 253 &  &  \\
17374.. & 435: &  &  & Yes &  \\
17435.. & 667$^a$ &  & & Yes &  \\
17505.. &  & 43.4,49.1 & 934 & & Yes? \\
17562.. & 1339 & 88: &  &  \\
18002.. & 792 & 35 & 408:
& Yes$^{b}$ &  \\
18075.. & 791 & 61 & & Yes &  Yes:?\\
18155.. & 890 &  & 472, 81: &  Yes &  \\
 &  &  &  &  \\
 \textit{Possible} &  &  &  &   \\
06503.. &  &  &  & &  \\
10532.. &  &  &  &  \\
12283.. & &  & 409: &  &  \\
16592.. &  &  & 444:: & & \\
17210.. & 506 & 33 &  &  \\
17370.. & 785: &  &  & Yes? & Yes:?\\
\hline
\end{tabular}

 \medskip
    \begin{minipage}{\linewidth}
    \textbf{Notes.} $^{a}$Value derived from the minima rather than from the period analysis, which yields a slightly longer period of 701 days.
$^{b}$Detected on the DASCH photographic plates; see the text for further details.
    \end{minipage}
    
\end{table}

\section{Individual objects}\label{sec:individual}
In this section, we summarize the information on the noteworthy objects from the SALT sample.

\subsection{Bona-fide symbiotics}

\noindent\textbf{2MASS~J14031865-5809349}\\
Prior to its confirmation as an S-type symbiotic binary in \citetalias{2014MNRAS.440.1410M}, this object was long known as an emission-line star \citep{1966PhDT.........3W}. Its quiescent spectrum reveals the presence of an M4 giant and a tentative detection of [\ion{Fe}{vii}], along with strong \ion{He}{ii} emission lines. 

 \begin{figure}
\centering
 \includegraphics[width=1\columnwidth]{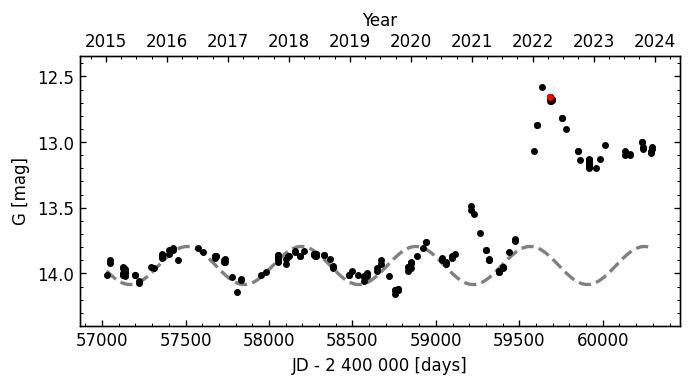}
 \caption{ \textit{Gaia} $G$-band light curve of 2MASS~J14031865-5809349. The time of the issued \textit{Gaia} Science Alert is marked with a red symbol. The underlying quiescent sinusoidal variability, with a period of 685 days, is shown in gray (see text for details).}
 \label{fig:outburst_14031_phot}
 \end{figure}

  \begin{figure}
\centering
 \includegraphics[width=1\columnwidth]{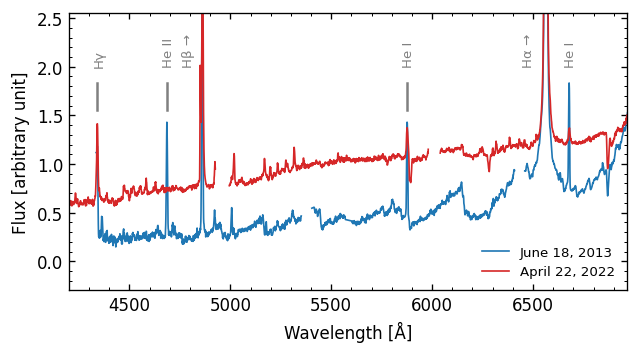}
 \caption{Comparison of the quiescent (blue) and outburst (red) spectra of 2MASS~J14031865-5809349, obtained with SALT. Selected emission lines are identified in gray. Most of the remaining, unmarked emission features, particularly in the outburst spectrum, are due to \ion{Fe}{ii}. }
 \label{fig:outburst_14031_spectrum}
 \end{figure}

 An outburst of this object was detected by the \textit{Gaia} satellite and reported as Gaia22bou on April 12, 2022 \citep{2022TNSTR.993....1H}, through the \textit{Gaia} Science Alerts (GSA) pipeline \citep[][]{2021A&A...652A..76H}. The outburst is clearly visible in our data, ASAS-SN and ATLAS photometric data (see Fig.~\ref{fig:photometry8}), and it is evident that the system had already been in an active state for several months prior to the alert. The \textit{Gaia} $G$-band light curve, available from the GSA website and shown in Fig.~\ref{fig:outburst_14031_phot}, reveals that the main outburst was preceded by a smaller brightening of approximately 0.5 mag in $G$, peaking around December 2020 (JD 2\,459\,210). This earlier event is also visible in our data (in particular in the $I$ filter), but is hidden in the noise in the ASAS-SN $g$-band data. Furthermore, pre-outburst activity likely began between July and September 2019, with the variability of the system deviating from the previously observed sinusoidal pattern, in particular well visible in \textit{Gaia} light curve.

 The period of the quiescent variability is $\sim$685 days (inferred from the \textit{Gaia G} light curve, but consistent with our $V$ data--see the maxima in the light curve in Fig. \ref{fig:photometry8} around JD 2\,457\,500 and JD 2\,458\,200), which we interpret as the orbital period of the binary. While other datasets suggest a possible period of 563 days (see Table \ref{tab:variability_paperI}), this shorter period is likely related to the morphology of the outburst phase (several maxima repeated with timescale close to this value) rather than orbital motion.

 Our data in the $V$ and $I$ filters indicate an outburst amplitude of around 2.5 and 1.0 mag, respectively. An amplitude of about 1.5 mag is observed in the broader \textit{Gaia} $G$-band. Several later rebrightenings are apparent in our data as well as the ASAS-SN and ATLAS light curves, and such photometric behaviour is typical of classical symbiotic star outbursts.

Following the \textit{Gaia} alert, two low-resolution spectra were obtained on April 18 and 20, 2022 (JD 2\,459\,687.9 and JD 2\,459\,689.7), and analyzed by \citet{2022ATel15340....1M}, who concluded that the star was undergoing a classical Z~And-type outburst, the first ever recorded for this system. During the outburst, high-ionization lines such as \ion{He}{ii} and [\ion{Fe}{vii}] disappeared from the spectra. We obtained an additional outburst spectrum with SALT on April 22, 2022 (JD 2\,459\,691.5). A comparison of the quiescent and outburst spectra is shown in Fig.~\ref{fig:outburst_14031_spectrum}. These data confirm a drop in the ionization temperature of the hot component: high-ionization lines such as [\ion{Fe}{vii}], \ion{He}{ii}, and [\ion{O}{iii}] are absent, while numerous \ion{Fe}{ii} lines emerge during outburst. Simultaneously, a rise in the blue continuum veiled the molecular bands of the cool companion.\\

\begin{figure}
\centering
 \includegraphics[width=1\columnwidth]{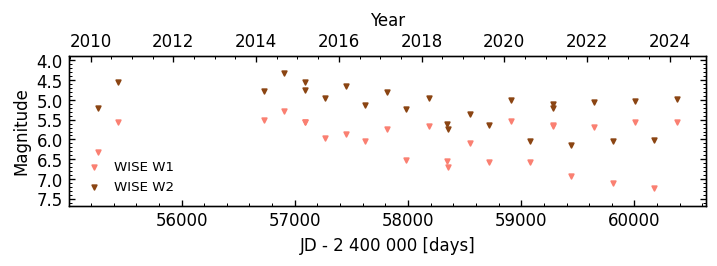}
 \caption{AllWISE and NEOWISE \textit{W1} and \textit{W2}-band light curves of 2MASS~J16422739-4133105.}
 \label{fig:16422_wise}
 \end{figure}

\noindent\textbf{2MASS J16422739-4133105}\\
The optical spectrum presented in \citetalias{2014MNRAS.440.1410M} allowed this target to be classified as a D-type symbiotic star, showing a rich emission-line spectrum, including Raman-scattered \ion{O}{vi} lines. The OGLE-IV, ATLAS, and our own data reveal large-amplitude variability with a period of about 369 days (Fig.~\ref{fig:photometry8}), consistent with a Mira pulsator, as expected for a D-type symbiotic system. The variability amplitude, however, appears to be strongly variable. Significant changes in amplitude (up to $\sim$1.5 mag) are also present in the mid-IR photometry from the Wide-field Infrared Survey Explorer \citep[WISE;][]{2010AJ....140.1868W}, both in the AllWISE release and in the NEOWISE measurements \citep{2011ApJ...731...53M,2014ApJ...792...30M}, see Fig. \ref{fig:16422_wise}. Owing to the $\sim$6-month cadence of NEOWISE and the variability period being close to one year, it is difficult to trace amplitude variations in detail. Nonetheless, the data suggest that when the system was slightly brighter in the mid-IR, both the optical and mid-IR amplitudes were smaller, likely related to dust production by the Mira and the resulting increased extinction.\\

\noindent\textbf{2MASS~J17160302-3322285}\\
The variability of this newly confirmed S-type symbiotic star was first reported by \citet{1997A&AS..123..507T}, where it is listed as Terz~V~4021 with an R-band magnitude between 15.8 and 17.6 mag. ATLAS $c$- and $o$-band data span about 3400~days and show clear variability (Fig.~\ref{fig:photometry}); however, period analysis is highly uncertain, particularly due to three distinct brightenings observed around JD 2\,458\,360, 2\,459\,750, and 2\,460\,800. The first two events had amplitudes of about 0.5 mag in $o$, while the third reached roughly 1.0 mag in $o$ and 1.5 mag in $c$. These amplitudes should be treated with caution, as the star lies in a relatively crowded field. OGLE-IV data suggest a shorter period of about 142 days, which we interpret as due to pulsations of the giant.\\

\noindent\textbf{2MASS~J17334728-2719266}\\
This star was classified as an S-type symbiotic system with an M2 giant donor and a typical symbiotic spectrum, including emission lines up to \ion{He}{ii}, in \citetalias{2014MNRAS.440.1410M}. While analyzing its variability here, we noticed a peculiar beating pattern and recovered two close periods of 161 and 179 days (Fig.~\ref{fig:photometry8}). Given the intriguing nature of this variability, we carried out a more detailed analysis, combining these data with additional infrared spectroscopy and imaging, and presented the results separately in \citet{2025arXiv251122988M}. The main conclusion of that study is that the beating pattern in the light curve is caused by the presence of two pulsating stars in close proximity. Moreover, the original identification of the symbiotic star was incorrect: it had been associated with the wrong \textit{Gaia} source. The object now confirmed as the symbiotic star exhibits a pulsation period of 161 days.\\ 

\noindent\textbf{2MASS~J17374702-2501120}\\
This star is classified here as a new S-type symbiotic system with an M3 cool component, exhibiting highly ionized emission lines such as \ion{He}{ii}, [\ion{Fe}{vii}], and \ion{O}{vi}.

In the photometric data (see Fig.~\ref{fig:photometry}), we identified several outbursts. Two brightenings are clearly visible in the OGLE-IV light curve, with maxima in April 2015 (JD 2\,457\,115) and June 2016 (JD 2\,457\,550), each reaching an amplitude of about 1 mag in the $I$ band. The second event is also detected in the ATLAS data, with a similar amplitude of $\sim$1 mag in the $o$ band. Two additional outbursts are captured by ATLAS: one in August 2017 (JD 2\,457\,975) with amplitudes of 2 mag in $o$ and 2.8 mag in $c$, and a less prominent one in July 2020 (JD 2\,459\,050), with amplitudes of about 1 mag in $o$ and 1.5 mag in $c$. The latter was also detected by ZTF. The recurrence timescale, amplitude, and duration of these events suggest that they are classical Z And-type symbiotic outbursts. No further outbursts have been observed since then.

Given the data quality, we used only the pre-outburst OGLE observations to determine the variability period, which we estimate to be approximately 435 days. We interpret this as the orbital period of the system.  \\

\noindent\textbf{2MASS~J17391715-3546593}\\
\citet{2014MNRAS.440.1410M} classified this object as an S-type symbiotic system, with a typical spectrum showing highly ionized emission lines, such as [\ion{Fe}{vii}] and \ion{O}{vi}, superimposed on the continuum of an early M-type giant. Previously, it had been considered a PN following its discovery by \citet{1994AN....315..235K}.

 \begin{figure}
\centering
 \includegraphics[width=1\columnwidth]{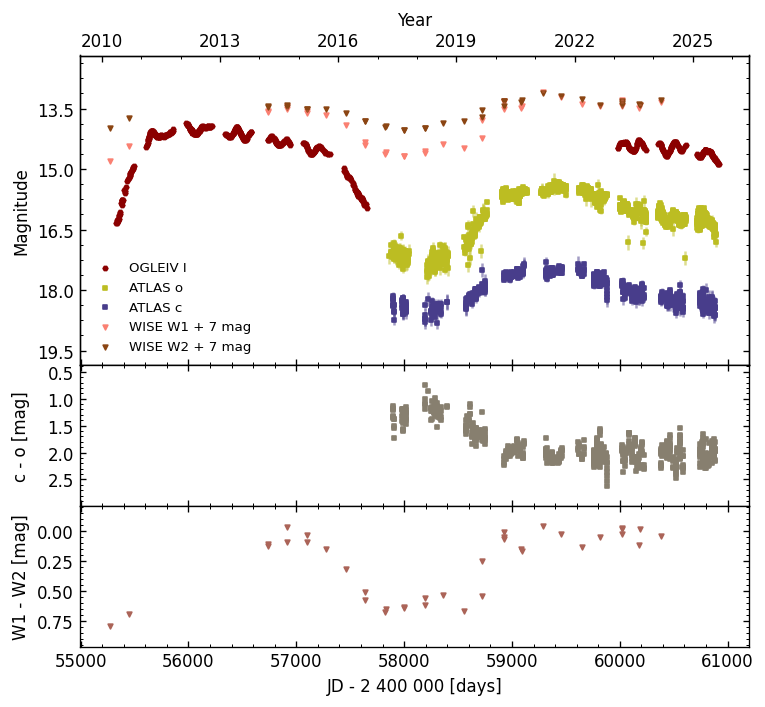}
 \caption{Photometric evolution of 2MASS~J17391715-3546593 over the past $\sim$16 years. The upper panel shows the OGLE, ATLAS, and WISE light curves of the object. The middle panel displays the ATLAS $c - o$ color evolution, while the bottom panel shows the WISE \textit{W1}$ - $\textit{W2} color evolution.}
 \label{fig:evolution_17391}
 \end{figure}

 While the spectrum is not unusual for a symbiotic star, the photometric evolution is highly peculiar (Fig.~\ref{fig:evolution_17391}). The OGLE-IV $I$-band light curve reveals a sharp brightening of at least 2.5 mag in 2010--2011, followed by five years of gradual decline accompanied by low-amplitude variability with a $\sim$153-day period. In 2016, the brightness dropped relatively abruptly, but OGLE coverage resumed only in 2023 with a single point consistent with another high state. This gap is well covered by ATLAS, which recorded the system in a low state and then documented the onset of a second high state in 2018--2019. Although a direct comparison between the two high states is not possible due to differences in photometric systems, a rough estimate suggests the brightening recurred after about $\sim$3\,200 days, with the most recent high state showing a similar slow decline that is now slightly longer than the previous one.

Part of the pre-2010 low state is covered by mid-IR photometry from WISE in the AllWISE release, while subsequent \textit{W1} and \textit{W2} measurements from the NEOWISE mission span the first high state, the following low state, and the second high state. The mid-IR light curves broadly mirror the optical behavior.

The color evolution is interesting. In the ATLAS optical bands ($c$–$o$), the source became bluer during the optical minimum, whereas in the mid-IR (\textit{W1}–\textit{W2}) it became redder. Such behavior is inconsistent with simple orbital variability (the amplitude-color dependence does not match) and is also unlike typical outburst activity, where the star becomes bluer as it brightens. 

A more plausible explanation is a dust-obscuration event in which direct light from the cool giant is attenuated while the emission from the nebula and/or hot component remains largely unaffected, leading to a bluer optical color. The reddening of the mid-IR SED could result either from the appearance of a warm dust component peaking more strongly in \textit{W2} than in \textit{W1}, or from suppression of the cooler stellar continuum while the other components of the system contribute little at these wavelengths.

A similar behavior has been reported in other symbiotic systems, particularly those hosting Mira donors. The best-studied case is RX Pup \citep[][see also Sect. 3 of \citealp{2000ASPC..199..431M}]{1999MNRAS.305..190M}. Analogous evolution is also well known in R~CrB stars \citep[e.g.,][]{1996PASP..108..225C,2025MNRAS.537.2635C}, and a comparable mechanism has been invoked to explain the "Great Dimming" of Betelgeuse in 2019--2020 \citep[e.g.,][]{2021A&A...650L..17K,2022ApJ...936...18D,2024A&A...685A.124J}.\\

\noindent\textbf{2MASS~J17435611-2506254}\\
Photometric data of this newly confirmed S-type symbiotic system show periodic variability with a period of 660 days, with a light curve morphology reminiscent of eclipses (Fig.~\ref{fig:photometry}). In addition to this variability, at least four outbursts with an amplitude >0.5 mag in $I$-band are clearly visible in the OGLE-IV light curve: in September 2010 ($\sim$JD 2\,455\,460), February 2014 (JD 2\,456\,710), February 2016 (JD 2\,457\,440), and August 2017 (JD 2\,457\,980; with the decline missing due to a seasonal gap). The last event is also covered by ATLAS data.

\begin{figure}
\centering
 \includegraphics[width=1\columnwidth]{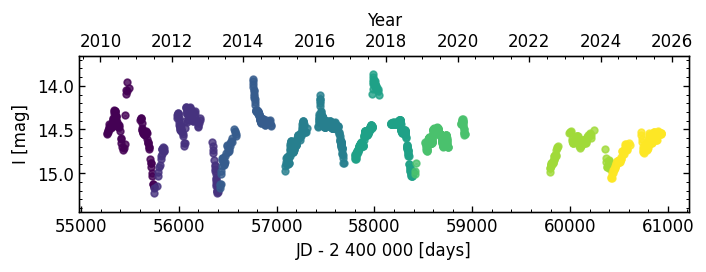}
 \includegraphics[width=1\columnwidth]{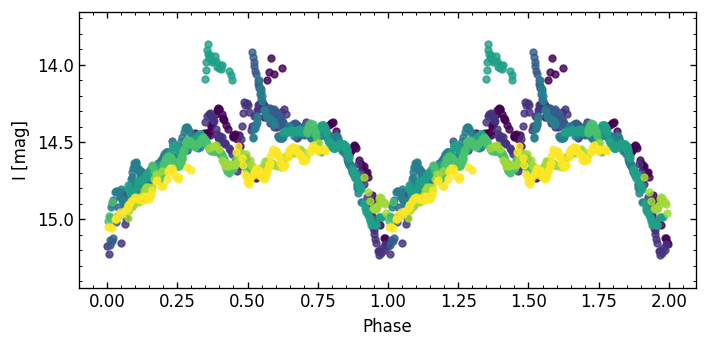}
 \caption{OGLE-IV $I$-band light curve of 2MASS~J17435611-2506254. The upper panel shows the full light curve, color-coded by cycle, the lower panel displays the phased light curve using the same color scheme.}
 \label{fig:ogle_17435}
 \end{figure}

The phased light curve (Fig.\ref{fig:ogle_17435}) indicates that all brightening events occur at a similar orbital phase. Its overall shape closely resembles that of FN Sgr, a symbiotic system hosting a magnetic white dwarf \citep{2023A&A...675A.140M}. In that case, the most plausible explanation involved enhanced visibility of the stream–disc overflow. Notably, in at least three cycles of 2MASS J17435611-2506254, a secondary dip around phase 0.4–0.5 mag is also present. This feature further strengthens the analogy with FN Sgr, where similar minima have been attributed to ellipsoidal variability.\\

\noindent\textbf{2MASS~J17505978-3012473}\\
Only OGLE photometry is available for this newly classified S-type symbiotic system. The light curve exhibits peculiar deep, eclipse-like minima. Folding the OGLE-IV data yields a very clean periodicity of 934 days, with the deep minima recurring at a consistent phase during that part of the survey. However, the deep minimum seen in the earlier OGLE-III light curve does not fall at the same phase when extrapolated with this period (it occurred about 130 days later than expected; phase shift of 0.14). 

The stars in the vicinity of 2MASS~J17505978-3012473 do not exhibit comparable variations, making it unlikely that the observed behavior is an artifact. Even disregarding the sudden change in periodicity, the amplitude of variability (noting that OGLE data are obtained in the $I$ band) is too large to be explained by eclipses in symbiotic systems. Deep eclipses are typically observed only during outbursts, when the hot companion dominates the optical flux and the red giant contributes minimally (see, for example, CI Cyg, \citealp{1991AJ....101..637K}; V618 Sgr, \citealp{2023MNRAS.523..163M}; Hen 3-860, \citealp{2022MNRAS.510.1404M}; or PU Vul, \citealp{2011ApJ...727...72K}). This is clearly not the case here, as the SALT spectrum obtained in 2014 shows both the clear signature of the red giant and the presence of highly ionized emission lines, typical for quiescent spectra of symbiotics (Fig. \ref{fig:spectra_1}). The unusually deep features in the light curve therefore remain without a satisfactory explanation and warrant further investigation.

In addition to the long-term variability, the system also displays short-term variations, most likely attributable to pulsations of the cool component. Period analysis of the OGLE-IV light curve reveals two significant periodicities (after prewhitening the data to remove the dominant periodic signal), at approximately 43 and 49 days, while the OGLE-III data show evidence only for the longer of these periods.\\

\noindent\textbf{2MASS~J18002542-1126324}\\
This newly discovered S-type symbiotic system was previously classified as a semi-regular variable with a period of 250 days based on ZTF data \citep{2020ApJS..249...18C}. However, our analysis shows that this is a yearly alias of a longer, 792-day period, which is consistently present across all datasets (Fig.~\ref{fig:photometry}).

 \begin{figure}
\centering
 \includegraphics[width=1\columnwidth]{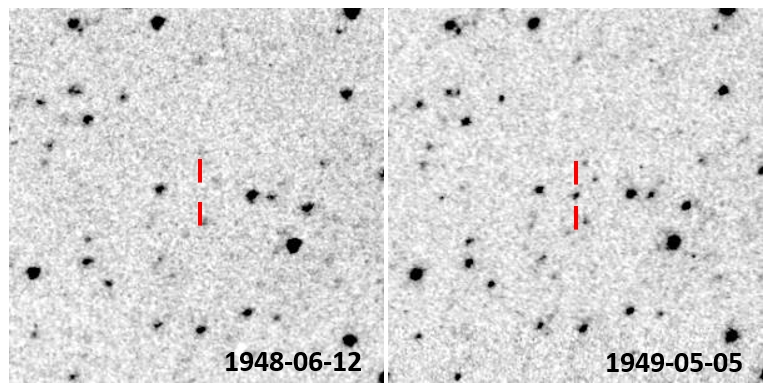}
 \caption{Two DASCH plates obtained before and during the 1949 outburst of 2MASS~J18002542-1126324.}
 \label{fig:dasch}
 \end{figure}

Furthermore, our analysis revealed that the star underwent an outburst in 1949, which has not been previously reported in the literature but is clearly visible in DASCH plate data (Fig.~\ref{fig:dasch}). The star is distinctly detected on 11 plates taken between May 3 and June 15, 1949, with magnitudes between 15.0 and 15.4. Following this, only a few plates from 1950--1951 have limiting magnitudes around 16. Most plates taken until 1953 have limiting magnitudes between 9 and 13. A gap in observations exists between 1953 and 1964. All plates from 1964 onward have limiting magnitudes between 10.0 and 15.0, fainter than the maximum brightness reached during the 1949 event. Notably, pre-outburst plates show no detection of the star down to a limiting magnitude of $\sim$17.5. The last such deep plate with a limiting magnitude greater than 15 mag was obtained in July 1948. Given the limited information, it is difficult to determine the nature of the outburst, whether it was a classical Z And-type symbiotic outburst or an eruption of a slow symbiotic nova.\\

\noindent\textbf{2MASS~J18075073-2516427}\\
This newly confirmed S-type symbiotic star, with an M3 cool component and a relatively hot companion, as evidenced by the presence of \ion{O}{vi} lines in the optical spectrum, experienced an outburst in June 2023 (around JD 2\,460\,100), clearly visible in ZTF data (Fig.~\ref{fig:photometry}). The event had an amplitude of approximately 1.1 mag in the $r$ band and 1.5 mag in the $g$ band, and is also detectable in the ATLAS light curve. Another, even more prominent outburst, seen in the ATLAS and OGLE-IV data, peaked around March 2025 (JD 2\,460\,750). Its amplitude was at least 1.5-2.0 mag in the ATLAS filters and 0.6 mag in \textit{I} band; however, the quiescent magnitude in the ATLAS filters is not well constrained owing to the noisy data. The observed evolution, duration, and recurrence timescale of these events are consistent with classical symbiotic outbursts. Finally, we note that the OGLE quiescent light curve reveals regular eclipse-like minima with a period of $\sim$786 days, broadly consistent with the results of the period analysis.\\

\begin{figure}
\centering
 \includegraphics[width=1\columnwidth]{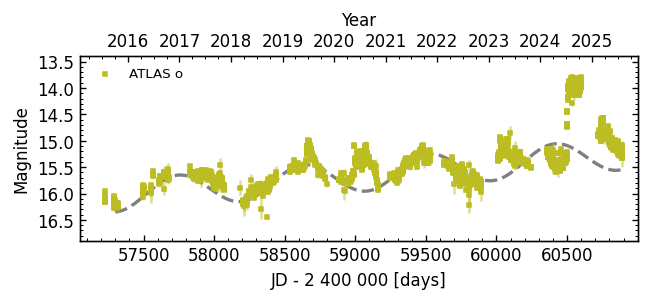}
 \hspace{10px}\includegraphics[width=0.495\columnwidth]{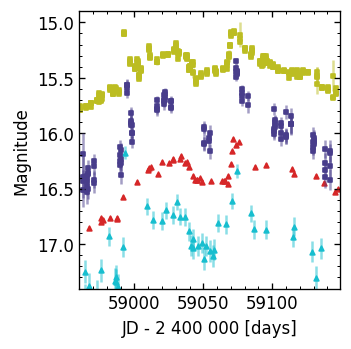}
  \includegraphics[width=0.495\columnwidth]{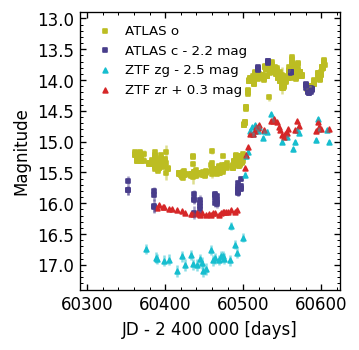}
 \caption{Variability of 2MASS~J18155753-0837357. The upper panel shows the ATLAS $o$-band light curve, fitted with a sinusoidal curve with a period of 890 days combined with a linear trend. The lower panels present details of the 2020 outburst and the outburst that has been ongoing since July 2024.}
 \label{fig:18155}
 \end{figure}

\noindent\textbf{2MASS~J18155753-0837357}\\
This is another newly confirmed S-type symbiotic star. Its photometric behavior is complex, with a recent outburst clearly visible in the ATLAS and ZTF light curves. Period analysis of the pre-outburst data suggested a period of about 472 days (Fig.~\ref{fig:photometry} and Table \ref{tab:variability}); however, this appears to be linked to additional activity rather than the orbital motion of the system. The light curve is more consistently modulated with a period of 890 days and shows a rising trend over the observed interval (Fig.~\ref{fig:18155}), which may correspond to the orbital period. 

Interestingly, the brightening that occurred around the minimum of the presumed sinusoidal variation in 2020 appears to have been abruptly "interrupted" for about 35 days (around JD 2\,459\,050). ZTF data suggest that the system became redder during this interval. Such a color change would be consistent with an eclipse of the hot component in a symbiotic system; however, the available data are insufficient to search for additional similar events. A comparable interruption of an outburst was observed in Gaia18aen \citep[][]{2020A&A...644A..49M}, where the source likewise reddened. Nevertheless, establishing a firm connection with orbital motion requires spectroscopic orbital data. Finally, we also note that part of the ATLAS $c$ light curve (between JD 2\,459\,650 and 2\,459\,860) shows hints of an 81-day period, which is, however, not detectable in other datasets.

\subsection{Possible symbiotic stars}

\noindent\textbf{2MASS~J06503882-0006394}\\
The absence of molecular bands in the spectrum, combined with the near-IR colors, suggests that this star is a possible D-type symbiotic system. It is listed as a semiregular variable with a period of $\sim$346 days in \citet{2020ApJS..249...18C}. The ATLAS and ZTF light curves (Fig.~\ref{fig:photometry5}) reveal complex photometric variability with no clear periodicity, characterized by several relatively deep minima that are not evenly spaced in time. If the object is indeed a D-type symbiotic star, its orbital period would likely be on the order of decades, and thus eclipses would not be expected to appear in the light curve.\\

\noindent\textbf{2MASS~J17370406-3324539}\\
The SALT/RSS optical spectrum of this candidate with a relatively low signal-to-noise ratio shows only Balmer lines and \ion{He}{i} (Fig.~\ref{fig:spectra_possible}). The red part of the spectrum exhibits possible molecular band features, suggesting the presence of a red giant. The ASAS-SN light curve is noisy and reveals no significant periodic variability (Fig.~\ref{fig:photometry5}). The OGLE-IV $I$-band data show a 0.3--0.4 mag increase in brightness at the beginning of the observed interval (JD < 2\,457\,000). Contemporaneous ASAS-SN $V$-band measurements display only a minor offset ($\sim$0.1 mag) from later observations, but the data quality is considerably lower. The $I$-band variability is consistent with a Z And-type outburst in a symbiotic system; however, the available data are insufficient to fully confirm this interpretation. Period analysis of the OGLE-IV light curve yields a period of $\sim$785 days, consistent with the minima in the data.\\

\begin{figure}
\centering
 \includegraphics[width=1\columnwidth]{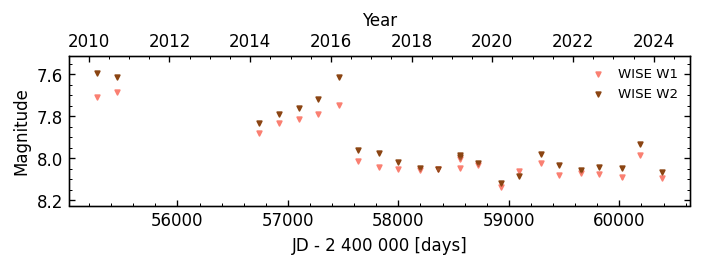}
 \caption{AllWISE and NEOWISE \textit{W1} and \textit{W2}-band light curves of 2MASS~J17460199-3303085.}
 \label{fig:17460_wise}
 \end{figure}

\noindent\textbf{2MASS~J17460199-3303085}\\
The ATLAS photometry is noisy but indicates periodic variability with a period of about 719 days, whereas our data suggest a slightly longer period of $\sim$775 days (Fig. \ref{fig:photometry10}, Table \ref{tab:variability_paperI}). In addition to the periodic signal, our light curve clearly shows what appears to be a decline from an outburst of more than 2 mag in the $V$ band. The same outburst is also visible in the NEOWISE \textit{W1} and \textit{W2} light curves, which exhibit a $\sim$0.5 mag brightening between 2014 and 2016 (Fig.~\ref{fig:17460_wise}). When combined with the AllWISE measurements in the same bands from 2010, the data suggest that the source maintained a comparable brightness level for at least six years, although possible dimming episodes between the AllWISE and NEOWISE epochs cannot be confirmed. Unfortunately, both our and the ATLAS data cover only the decline phase of this outburst.

We only have a lower limit on the amplitudes in the $V$ and $I$ bands, but they are consistent with typical Z And-type outbursts of symbiotic stars. The amplitudes in \textit{W1} and \textit{W2} are also in line with those observed during outbursts of other symbiotic systems \citep[e.g., V390 Sco;][]{2024AN....34540017M}. The SALT spectrum used for the tentative symbiotic classification \citepalias{2014MNRAS.440.1410M} was obtained in September 2013. If the system was in outburst at that time, it could explain why the spectrum shows only low-ionization emission lines (Balmer lines of \ion{H}{i}, \ion{He}{i}, and faint [\ion{O}{iii}]). A quiescent spectrum could therefore provide a definitive confirmation of the symbiotic nature of this target.\\

\section{Conclusions}\label{sec:conclusions}
In this work, we continued the systematic search for new Galactic southern symbiotic stars, selecting candidates from SHS and 2MASS data following the criteria of \citetalias{2014MNRAS.440.1410M}. Spectroscopy obtained with SALT was used to confirm the symbiotic nature of the selected candidates and to characterize the components of the entire sample, including those from \citetalias{2014MNRAS.440.1410M}.

We report the discovery of 14 new bona-fide symbiotic systems and 6 additional strong candidates. Of the newly confirmed systems, 13 are S-type with oxygen-rich M-type giants, and one is D-type. All show [\ion{Fe}{vii}] and/or \ion{O}{vi} emission lines in their optical spectra, with one system possibly also exhibiting [\ion{Fe}{x}]. Among the candidates, three are likely D-type and three S-type.

We analyzed the photometric variability of the systems, confirming most as variables, with the majority showing periodic modulation attributable to orbital motion or pulsations. In total, we present possible (photometric) orbital periods for 19 bona-fide and 3 possible symbiotic stars. These periods should be confirmed through dedicated spectroscopic follow-up.

For 2MASS~J17391715-3546593, the unusual photometric evolution is most likely caused by dust-obscuration events. Eight sources show evidence of outburst activity in their light curves. Notably, 2MASS~J17435611-2506254 exhibits several brightenings occurring at similar orbital phases, closely resembling evolution of FN Sgr, symbiotic binary with magnetic white dwarf.

\section*{Acknowledgements}

We thank the anonymous referee for the careful review and helpful suggestions that improved the manuscript. The research of JaM was supported by the Czech Science Foundation (GACR) project no. 24-10608O. JMik was supported by the Polish National Science Centre (NCN) grant
2023/48/Q/ST9/00138. KI was supported by the Polish National Science Centre (NCN) grant 2024/55/D/ST9/01713. The OGLE project has received funding from the Polish National Science
Centre grant OPUS-28 2024/55/B/ST9/00447 to AU. The paper is based on spectroscopic observations made with the Southern
African Large Telescope (SALT). Polish participation in SALT is funded by grant No. MEiN 2021/WK/01. 

\section*{Data Availability}
Our photometric data will be shared on reasonable request to the corresponding author. Additional photometric observations are accessible from the websites of the surveys. Spectra are available in the SALT archive.



\bibliographystyle{mnras}
\bibliography{example} 

\begin{thebibliography}{}
\makeatletter
\relax
\def\mn@urlcharsother{\let\do\@makeother \do\$\do\&\do\#\do\^\do\_\do\%\do\~}
\def\mn@doi{\begingroup\mn@urlcharsother \@ifnextchar [ {\mn@doi@}
  {\mn@doi@[]}}
\def\mn@doi@[#1]#2{\def\@tempa{#1}\ifx\@tempa\@empty \href
  {http://dx.doi.org/#2} {doi:#2}\else \href {http://dx.doi.org/#2} {#1}\fi
  \endgroup}
\def\mn@eprint#1#2{\mn@eprint@#1:#2::\@nil}
\def\mn@eprint@arXiv#1{\href {http://arxiv.org/abs/#1} {{\tt arXiv:#1}}}
\def\mn@eprint@dblp#1{\href {http://dblp.uni-trier.de/rec/bibtex/#1.xml}
  {dblp:#1}}
\def\mn@eprint@#1:#2:#3:#4\@nil{\def\@tempa {#1}\def\@tempb {#2}\def\@tempc
  {#3}\ifx \@tempc \@empty \let \@tempc \@tempb \let \@tempb \@tempa \fi \ifx
  \@tempb \@empty \def\@tempb {arXiv}\fi \@ifundefined
  {mn@eprint@\@tempb}{\@tempb:\@tempc}{\expandafter \expandafter \csname
  mn@eprint@\@tempb\endcsname \expandafter{\@tempc}}}

\bibitem[\protect\citeauthoryear{{Akras}}{{Akras}}{2023}]{2023MNRAS.519.6044A}
{Akras} S.,  2023, \mn@doi [\mnras] {10.1093/mnras/stad096}, \href
  {https://ui.adsabs.harvard.edu/abs/2023MNRAS.519.6044A} {519, 6044}

\bibitem[\protect\citeauthoryear{{Akras}, {Guzman-Ramirez}, {Leal-Ferreira}  \&
  {Ramos-Larios}}{{Akras} et~al.}{2019}]{2019ApJS..240...21A}
{Akras} S.,  {Guzman-Ramirez} L.,  {Leal-Ferreira} M.~L.,   {Ramos-Larios} G.,
  2019, \mn@doi [\apjs] {10.3847/1538-4365/aaf88c}, \href
  {https://ui.adsabs.harvard.edu/abs/2019ApJS..240...21A} {240, 21}

\bibitem[\protect\citeauthoryear{{Akras}, {Gon{\c{c}}alves}, {Alvarez-Candal}
  \& {Pereira}}{{Akras} et~al.}{2021}]{2021MNRAS.502.2513A}
{Akras} S.,  {Gon{\c{c}}alves} D.~R.,  {Alvarez-Candal} A.,   {Pereira} C.~B.,
  2021, \mn@doi [\mnras] {10.1093/mnras/stab195}, \href
  {https://ui.adsabs.harvard.edu/abs/2021MNRAS.502.2513A} {502, 2513}

\bibitem[\protect\citeauthoryear{{Allen}}{{Allen}}{1984}]{1984PASA....5..369A}
{Allen} D.~A.,  1984, \mn@doi [\pasa] {10.1017/S1323358000017215}, \href
  {https://ui.adsabs.harvard.edu/abs/1984PASA....5..369A} {5, 369}

\bibitem[\protect\citeauthoryear{{Astropy Collaboration} et~al.,}{{Astropy
  Collaboration} et~al.}{2013}]{2013A&A...558A..33A}
{Astropy Collaboration} et~al., 2013, \mn@doi [\aap]
  {10.1051/0004-6361/201322068}, \href
  {https://ui.adsabs.harvard.edu/abs/2013A&A...558A..33A} {558, A33}

\bibitem[\protect\citeauthoryear{{Astropy Collaboration} et~al.,}{{Astropy
  Collaboration} et~al.}{2018}]{2018AJ....156..123A}
{Astropy Collaboration} et~al., 2018, \mn@doi [\aj] {10.3847/1538-3881/aabc4f},
  \href {https://ui.adsabs.harvard.edu/abs/2018AJ....156..123A} {156, 123}

\bibitem[\protect\citeauthoryear{{Astropy Collaboration} et~al.,}{{Astropy
  Collaboration} et~al.}{2022}]{2022ApJ...935..167A}
{Astropy Collaboration} et~al., 2022, \mn@doi [\apj]
  {10.3847/1538-4357/ac7c74}, \href
  {https://ui.adsabs.harvard.edu/abs/2022ApJ...935..167A} {935, 167}

\bibitem[\protect\citeauthoryear{{Bailer-Jones}, {Rybizki}, {Fouesneau},
  {Demleitner}  \& {Andrae}}{{Bailer-Jones} et~al.}{2021}]{2021AJ....161..147B}
{Bailer-Jones} C.~A.~L.,  {Rybizki} J.,  {Fouesneau} M.,  {Demleitner} M.,
  {Andrae} R.,  2021, \mn@doi [\aj] {10.3847/1538-3881/abd806}, \href
  {https://ui.adsabs.harvard.edu/abs/2021AJ....161..147B} {161, 147}

\bibitem[\protect\citeauthoryear{{Ball}, {Bromley}  \& {Kenyon}}{{Ball}
  et~al.}{2025}]{2025arXiv250620505B}
{Ball} S.~E.,  {Bromley} B.~C.,   {Kenyon} S.~J.,  2025, \mn@doi [arXiv
  e-prints] {10.48550/arXiv.2506.20505}, \href
  {https://ui.adsabs.harvard.edu/abs/2025arXiv250620505B} {p. arXiv:2506.20505}

\bibitem[\protect\citeauthoryear{{Belczy{\'n}ski}, {Miko{\l}ajewska}, {Munari},
  {Ivison}  \& {Friedjung}}{{Belczy{\'n}ski}
  et~al.}{2000}]{2000A&AS..146..407B}
{Belczy{\'n}ski} K.,  {Miko{\l}ajewska} J.,  {Munari} U.,  {Ivison} R.~J.,
  {Friedjung} M.,  2000, \mn@doi [\aaps] {10.1051/aas:2000280}, \href
  {https://ui.adsabs.harvard.edu/abs/2000A&AS..146..407B} {146, 407}

\bibitem[\protect\citeauthoryear{{Botello}, {Sabin}  \&
  {G{\'o}mez-Mu{\~n}oz}}{{Botello} et~al.}{2025}]{2025MNRAS.tmp.1318B}
{Botello} M.~K.,  {Sabin} L.,   {G{\'o}mez-Mu{\~n}oz} M.~A.,  2025, \mn@doi
  [\mnras] {10.1093/mnras/staf1342}, \href
  {https://ui.adsabs.harvard.edu/abs/2025MNRAS.tmp.1318B} {}

\bibitem[\protect\citeauthoryear{{Bovy}, {Rix}, {Green}, {Schlafly}  \&
  {Finkbeiner}}{{Bovy} et~al.}{2016}]{2016ApJ...818..130B}
{Bovy} J.,  {Rix} H.-W.,  {Green} G.~M.,  {Schlafly} E.~F.,   {Finkbeiner}
  D.~P.,  2016, \mn@doi [\apj] {10.3847/0004-637X/818/2/130}, \href
  {https://ui.adsabs.harvard.edu/abs/2016ApJ...818..130B} {818, 130}

\bibitem[\protect\citeauthoryear{{Buckley}, {Swart}  \& {Meiring}}{{Buckley}
  et~al.}{2006}]{2006SPIE.6267E..0ZB}
{Buckley} D. A.~H.,  {Swart} G.~P.,   {Meiring} J.~G.,  2006, in {Stepp} L.~M.,
   ed.,  Society of Photo-Optical Instrumentation Engineers (SPIE) Conference
  Series Vol. 6267, Ground-based and Airborne Telescopes. p. 62670Z,
  \mn@doi{10.1117/12.673750}

\bibitem[\protect\citeauthoryear{{Burgh}, {Nordsieck}, {Kobulnicky},
  {Williams}, {O'Donoghue}, {Smith}  \& {Percival}}{{Burgh}
  et~al.}{2003}]{2003SPIE.4841.1463B}
{Burgh} E.~B.,  {Nordsieck} K.~H.,  {Kobulnicky} H.~A.,  {Williams} T.~B.,
  {O'Donoghue} D.,  {Smith} M.~P.,   {Percival} J.~W.,  2003, in {Iye} M.,
  {Moorwood} A. F.~M.,  eds,  Society of Photo-Optical Instrumentation
  Engineers (SPIE) Conference Series Vol. 4841, Instrument Design and
  Performance for Optical/Infrared Ground-based Telescopes. pp 1463--1471,
  \mn@doi{10.1117/12.460312}

\bibitem[\protect\citeauthoryear{{Cardelli}, {Clayton}  \& {Mathis}}{{Cardelli}
  et~al.}{1989}]{1989ApJ...345..245C}
{Cardelli} J.~A.,  {Clayton} G.~C.,   {Mathis} J.~S.,  1989, \mn@doi [\apj]
  {10.1086/167900}, \href
  {https://ui.adsabs.harvard.edu/abs/1989ApJ...345..245C} {345, 245}

\bibitem[\protect\citeauthoryear{{Chen}, {Wang}, {Deng}, {de Grijs}, {Yang}  \&
  {Tian}}{{Chen} et~al.}{2020}]{2020ApJS..249...18C}
{Chen} X.,  {Wang} S.,  {Deng} L.,  {de Grijs} R.,  {Yang} M.,   {Tian} H.,
  2020, \mn@doi [\apjs] {10.3847/1538-4365/ab9cae}, \href
  {https://ui.adsabs.harvard.edu/abs/2020ApJS..249...18C} {249, 18}

\bibitem[\protect\citeauthoryear{{Chen}, {Wang}, {Li}, {Ma}, {Luo}, {Zhang},
  {Ding}  \& {Zhang}}{{Chen} et~al.}{2025}]{2025ApJ...987..147C}
{Chen} J.,  {Wang} L.,  {Li} Y.-B.,  {Ma} X.-X.,  {Luo} A.~L.,  {Zhang} Z.-C.,
  {Ding} M.-Y.,   {Zhang} K.,  2025, \mn@doi [\apj] {10.3847/1538-4357/addec3},
  \href {https://ui.adsabs.harvard.edu/abs/2025ApJ...987..147C} {987, 147}

\bibitem[\protect\citeauthoryear{{Clayton}}{{Clayton}}{1996}]{1996PASP..108..225C}
{Clayton} G.~C.,  1996, \mn@doi [\pasp] {10.1086/133715}, \href
  {https://ui.adsabs.harvard.edu/abs/1996PASP..108..225C} {108, 225}

\bibitem[\protect\citeauthoryear{{Corradi} et~al.,}{{Corradi}
  et~al.}{2008}]{2008A&A...480..409C}
{Corradi} R.~L.~M.,  et~al., 2008, \mn@doi [\aap] {10.1051/0004-6361:20078989},
  \href {https://ui.adsabs.harvard.edu/abs/2008A&A...480..409C} {480, 409}

\bibitem[\protect\citeauthoryear{{Corradi} et~al.,}{{Corradi}
  et~al.}{2010}]{2010A&A...509A..41C}
{Corradi} R.~L.~M.,  et~al., 2010, \mn@doi [\aap]
  {10.1051/0004-6361/200913231}, \href
  {https://ui.adsabs.harvard.edu/abs/2010A&A...509A..41C} {509, A41}

\bibitem[\protect\citeauthoryear{{Crawford}, {Soon}, {Clayton}, {Tisserand},
  {Bedding}, {Clark}  \& {Lee}}{{Crawford} et~al.}{2025}]{2025MNRAS.537.2635C}
{Crawford} C.~L.,  {Soon} J.,  {Clayton} G.~C.,  {Tisserand} P.,  {Bedding}
  T.~R.,  {Clark} C.~J.,   {Lee} C.-U.,  2025, \mn@doi [\mnras]
  {10.1093/mnras/staf215}, \href
  {https://ui.adsabs.harvard.edu/abs/2025MNRAS.537.2635C} {537, 2635}

\bibitem[\protect\citeauthoryear{{Drew} et~al.,}{{Drew}
  et~al.}{2005}]{2005MNRAS.362..753D}
{Drew} J.~E.,  et~al., 2005, \mn@doi [\mnras]
  {10.1111/j.1365-2966.2005.09330.x}, \href
  {https://ui.adsabs.harvard.edu/abs/2005MNRAS.362..753D} {362, 753}

\bibitem[\protect\citeauthoryear{{Drimmel}, {Cabrera-Lavers}  \&
  {L{\'o}pez-Corredoira}}{{Drimmel} et~al.}{2003}]{2003A&A...409..205D}
{Drimmel} R.,  {Cabrera-Lavers} A.,   {L{\'o}pez-Corredoira} M.,  2003, \mn@doi
  [\aap] {10.1051/0004-6361:20031070}, \href
  {https://ui.adsabs.harvard.edu/abs/2003A&A...409..205D} {409, 205}

\bibitem[\protect\citeauthoryear{{Dupree} et~al.,}{{Dupree}
  et~al.}{2022}]{2022ApJ...936...18D}
{Dupree} A.~K.,  et~al., 2022, \mn@doi [\apj] {10.3847/1538-4357/ac7853}, \href
  {https://ui.adsabs.harvard.edu/abs/2022ApJ...936...18D} {936, 18}

\bibitem[\protect\citeauthoryear{{Eyer} et~al.,}{{Eyer}
  et~al.}{2023}]{2023A&A...674A..13E}
{Eyer} L.,  et~al., 2023, \mn@doi [\aap] {10.1051/0004-6361/202244242}, \href
  {https://ui.adsabs.harvard.edu/abs/2023A&A...674A..13E} {674, A13}

\bibitem[\protect\citeauthoryear{{Gaia Collaboration} et~al.,}{{Gaia
  Collaboration} et~al.}{2023}]{2023A&A...674A...1G}
{Gaia Collaboration} et~al., 2023, \mn@doi [\aap]
  {10.1051/0004-6361/202243940}, \href
  {https://ui.adsabs.harvard.edu/abs/2023A&A...674A...1G} {674, A1}

\bibitem[\protect\citeauthoryear{{Green}, {Schlafly}, {Zucker}, {Speagle}  \&
  {Finkbeiner}}{{Green} et~al.}{2019}]{2019ApJ...887...93G}
{Green} G.~M.,  {Schlafly} E.,  {Zucker} C.,  {Speagle} J.~S.,   {Finkbeiner}
  D.,  2019, \mn@doi [\apj] {10.3847/1538-4357/ab5362}, \href
  {https://ui.adsabs.harvard.edu/abs/2019ApJ...887...93G} {887, 93}

\bibitem[\protect\citeauthoryear{{Gromadzki}, {Miko{\l}ajewska}  \&
  {Soszy{\'n}ski}}{{Gromadzki} et~al.}{2013}]{2013AcA....63..405G}
{Gromadzki} M.,  {Miko{\l}ajewska} J.,   {Soszy{\'n}ski} I.,  2013, \mn@doi
  [\actaa] {10.48550/arXiv.1312.6063}, \href
  {https://ui.adsabs.harvard.edu/abs/2013AcA....63..405G} {63, 405}

\bibitem[\protect\citeauthoryear{{Gutierrez-Moreno}, {Moreno}  \&
  {Cortes}}{{Gutierrez-Moreno} et~al.}{1995}]{1995PASP..107..462G}
{Gutierrez-Moreno} A.,  {Moreno} H.,   {Cortes} G.,  1995, \mn@doi [\pasp]
  {10.1086/133575}, \href
  {https://ui.adsabs.harvard.edu/abs/1995PASP..107..462G} {107, 462}

\bibitem[\protect\citeauthoryear{{Hodgkin} et~al.,}{{Hodgkin}
  et~al.}{2021}]{2021A&A...652A..76H}
{Hodgkin} S.~T.,  et~al., 2021, \mn@doi [\aap] {10.1051/0004-6361/202140735},
  \href {https://ui.adsabs.harvard.edu/abs/2021A&A...652A..76H} {652, A76}

\bibitem[\protect\citeauthoryear{{Hodgkin} et~al.,}{{Hodgkin}
  et~al.}{2022}]{2022TNSTR.993....1H}
{Hodgkin} S.~T.,  et~al., 2022, Transient Name Server Discovery Report, \href
  {https://ui.adsabs.harvard.edu/abs/2022TNSTR.993....1H} {2022-993, 1}

\bibitem[\protect\citeauthoryear{{Iijima}}{{Iijima}}{1981}]{1981ASIC...69..517I}
{Iijima} T.,  1981, in Photometric and Spectroscopic Binary Systems. p.~517,
  \mn@doi{10.1007/978-94-009-8486-8_27}

\bibitem[\protect\citeauthoryear{{I{\l}kiewicz} \&
  {Miko{\l}ajewska}}{{I{\l}kiewicz} \&
  {Miko{\l}ajewska}}{2017}]{2017A&A...606A.110I}
{I{\l}kiewicz} K.,  {Miko{\l}ajewska} J.,  2017, \mn@doi [\aap]
  {10.1051/0004-6361/201731497}, \href
  {https://ui.adsabs.harvard.edu/abs/2017A&A...606A.110I} {606, A110}

\bibitem[\protect\citeauthoryear{{Jadlovsk{\'y}} et~al.,}{{Jadlovsk{\'y}}
  et~al.}{2024}]{2024A&A...685A.124J}
{Jadlovsk{\'y}} D.,  et~al., 2024, \mn@doi [\aap]
  {10.1051/0004-6361/202348846}, \href
  {https://ui.adsabs.harvard.edu/abs/2024A&A...685A.124J} {685, A124}

\bibitem[\protect\citeauthoryear{{Kato}, {Hachisu}, {Cassatella}  \&
  {Gonz{\'a}lez-Riestra}}{{Kato} et~al.}{2011}]{2011ApJ...727...72K}
{Kato} M.,  {Hachisu} I.,  {Cassatella} A.,   {Gonz{\'a}lez-Riestra} R.,  2011,
  \mn@doi [\apj] {10.1088/0004-637X/727/2/72}, \href
  {https://ui.adsabs.harvard.edu/abs/2011ApJ...727...72K} {727, 72}

\bibitem[\protect\citeauthoryear{{Kenyon} \& {Fernandez-Castro}}{{Kenyon} \&
  {Fernandez-Castro}}{1987}]{1987AJ.....93..938K}
{Kenyon} S.~J.,  {Fernandez-Castro} T.,  1987, \mn@doi [\aj] {10.1086/114379},
  \href {https://ui.adsabs.harvard.edu/abs/1987AJ.....93..938K} {93, 938}

\bibitem[\protect\citeauthoryear{{Kenyon}, {Oliversen}, {Miko{\l}ajewska},
  {Mikolajewski}, {Stencel}, {Garcia}  \& {Anderson}}{{Kenyon}
  et~al.}{1991}]{1991AJ....101..637K}
{Kenyon} S.~J.,  {Oliversen} N.~A.,  {Miko{\l}ajewska} J.,  {Mikolajewski} M.,
  {Stencel} R.~E.,  {Garcia} M.~R.,   {Anderson} C.~M.,  1991, \mn@doi [\aj]
  {10.1086/115712}, \href
  {https://ui.adsabs.harvard.edu/abs/1991AJ....101..637K} {101, 637}

\bibitem[\protect\citeauthoryear{{Kenyon}, {Livio}, {Miko{\l}ajewska}  \&
  {Tout}}{{Kenyon} et~al.}{1993}]{1993ApJ...407L..81K}
{Kenyon} S.~J.,  {Livio} M.,  {Miko{\l}ajewska} J.,   {Tout} C.~A.,  1993,
  \mn@doi [\apjl] {10.1086/186811}, \href
  {https://ui.adsabs.harvard.edu/abs/1993ApJ...407L..81K} {407, L81}

\bibitem[\protect\citeauthoryear{{Kobulnicky}, {Nordsieck}, {Burgh}, {Smith},
  {Percival}, {Williams}  \& {O'Donoghue}}{{Kobulnicky}
  et~al.}{2003}]{2003SPIE.4841.1634K}
{Kobulnicky} H.~A.,  {Nordsieck} K.~H.,  {Burgh} E.~B.,  {Smith} M.~P.,
  {Percival} J.~W.,  {Williams} T.~B.,   {O'Donoghue} D.,  2003, in {Iye} M.,
  {Moorwood} A. F.~M.,  eds,  Society of Photo-Optical Instrumentation
  Engineers (SPIE) Conference Series Vol. 4841, Instrument Design and
  Performance for Optical/Infrared Ground-based Telescopes. pp 1634--1644,
  \mn@doi{10.1117/12.460315}

\bibitem[\protect\citeauthoryear{{Kochanek} et~al.,}{{Kochanek}
  et~al.}{2017}]{2017PASP..129j4502K}
{Kochanek} C.~S.,  et~al., 2017, \mn@doi [\pasp] {10.1088/1538-3873/aa80d9},
  \href {https://ui.adsabs.harvard.edu/abs/2017PASP..129j4502K} {129, 104502}

\bibitem[\protect\citeauthoryear{{Kohoutek}}{{Kohoutek}}{1994}]{1994AN....315..235K}
{Kohoutek} L.,  1994, \mn@doi [Astronomische Nachrichten]
  {10.1002/asna.2103150305}, \href
  {https://ui.adsabs.harvard.edu/abs/1994AN....315..235K} {315, 235}

\bibitem[\protect\citeauthoryear{{Kohoutek} \& {Wehmeyer}}{{Kohoutek} \&
  {Wehmeyer}}{2003}]{2003AN....324..437K}
{Kohoutek} L.,  {Wehmeyer} R.,  2003, \mn@doi [Astronomische Nachrichten]
  {10.1002/asna.200310089}, \href
  {https://ui.adsabs.harvard.edu/abs/2003AN....324..437K} {324, 437}

\bibitem[\protect\citeauthoryear{{Kravchenko} et~al.,}{{Kravchenko}
  et~al.}{2021}]{2021A&A...650L..17K}
{Kravchenko} K.,  et~al., 2021, \mn@doi [\aap] {10.1051/0004-6361/202039801},
  \href {https://ui.adsabs.harvard.edu/abs/2021A&A...650L..17K} {650, L17}

\bibitem[\protect\citeauthoryear{{Laversveiler}, {Gon{\c{c}}alves},
  {Rocha-Pinto}  \& {Merc}}{{Laversveiler} et~al.}{2025}]{2025A&A...698A.155L}
{Laversveiler} M.,  {Gon{\c{c}}alves} D.~R.,  {Rocha-Pinto} H.~J.,   {Merc} J.,
   2025, \mn@doi [\aap] {10.1051/0004-6361/202451548}, \href
  {https://ui.adsabs.harvard.edu/abs/2025A&A...698A.155L} {698, A155}

\bibitem[\protect\citeauthoryear{{Laycock}, {Tang}, {Grindlay}, {Los}, {Simcoe}
   \& {Mink}}{{Laycock} et~al.}{2010}]{2010AJ....140.1062L}
{Laycock} S.,  {Tang} S.,  {Grindlay} J.,  {Los} E.,  {Simcoe} R.,   {Mink} D.,
   2010, \mn@doi [\aj] {10.1088/0004-6256/140/4/1062}, \href
  {https://ui.adsabs.harvard.edu/abs/2010AJ....140.1062L} {140, 1062}

\bibitem[\protect\citeauthoryear{{Lebzelter} et~al.,}{{Lebzelter}
  et~al.}{2023}]{2023A&A...674A..15L}
{Lebzelter} T.,  et~al., 2023, \mn@doi [\aap] {10.1051/0004-6361/202244241},
  \href {https://ui.adsabs.harvard.edu/abs/2023A&A...674A..15L} {674, A15}

\bibitem[\protect\citeauthoryear{{Leedj{\"a}rv}, {G{\'a}lis}, {Hric}, {Merc}
  \& {Burmeister}}{{Leedj{\"a}rv} et~al.}{2016}]{2016MNRAS.456.2558L}
{Leedj{\"a}rv} L.,  {G{\'a}lis} R.,  {Hric} L.,  {Merc} J.,   {Burmeister} M.,
  2016, \mn@doi [\mnras] {10.1093/mnras/stv2807}, \href
  {https://ui.adsabs.harvard.edu/abs/2016MNRAS.456.2558L} {456, 2558}

\bibitem[\protect\citeauthoryear{{Lomb}}{{Lomb}}{1976}]{1976Ap&SS..39..447L}
{Lomb} N.~R.,  1976, \mn@doi [\apss] {10.1007/BF00648343}, \href
  {https://ui.adsabs.harvard.edu/abs/1976Ap&SS..39..447L} {39, 447}

\bibitem[\protect\citeauthoryear{{Lucy} et~al.,}{{Lucy}
  et~al.}{2024}]{2024arXiv241200855L}
{Lucy} A.~B.,  et~al., 2024, \mn@doi [arXiv e-prints]
  {10.48550/arXiv.2412.00855}, \href
  {https://ui.adsabs.harvard.edu/abs/2024arXiv241200855L} {p. arXiv:2412.00855}

\bibitem[\protect\citeauthoryear{{Magdolen}, {Dobrotka}, {Orio},
  {Miko{\l}ajewska}, {Vanderburg}, {Monard}, {Aloisi}  \&
  {Bez{\'a}k}}{{Magdolen} et~al.}{2023}]{2023A&A...675A.140M}
{Magdolen} J.,  {Dobrotka} A.,  {Orio} M.,  {Miko{\l}ajewska} J.,  {Vanderburg}
  A.,  {Monard} B.,  {Aloisi} R.,   {Bez{\'a}k} P.,  2023, \mn@doi [\aap]
  {10.1051/0004-6361/202345935}, \href
  {https://ui.adsabs.harvard.edu/abs/2023A&A...675A.140M} {675, A140}

\bibitem[\protect\citeauthoryear{{Magrini}, {Corradi}  \& {Munari}}{{Magrini}
  et~al.}{2003}]{2003ASPC..303..539M}
{Magrini} L.,  {Corradi} R.~L.~M.,   {Munari} U.,  2003, in {Corradi} R.~L.~M.,
   {Miko{\l}ajewska} J.,   {Mahoney} T.~J.,  eds,  Astronomical Society of the
  Pacific Conference Series Vol. 303, Symbiotic Stars Probing Stellar
  Evolution. p.~539 (\mn@eprint {arXiv} {astro-ph/0208085}),
  \mn@doi{10.48550/arXiv.astro-ph/0208085}

\bibitem[\protect\citeauthoryear{{Mainzer} et~al.,}{{Mainzer}
  et~al.}{2011}]{2011ApJ...731...53M}
{Mainzer} A.,  et~al., 2011, \mn@doi [\apj] {10.1088/0004-637X/731/1/53}, \href
  {https://ui.adsabs.harvard.edu/abs/2011ApJ...731...53M} {731, 53}

\bibitem[\protect\citeauthoryear{{Mainzer} et~al.,}{{Mainzer}
  et~al.}{2014}]{2014ApJ...792...30M}
{Mainzer} A.,  et~al., 2014, \mn@doi [\apj] {10.1088/0004-637X/792/1/30}, \href
  {https://ui.adsabs.harvard.edu/abs/2014ApJ...792...30M} {792, 30}

\bibitem[\protect\citeauthoryear{{Marshall}, {Robin}, {Reyl{\'e}}, {Schultheis}
   \& {Picaud}}{{Marshall} et~al.}{2006}]{2006A&A...453..635M}
{Marshall} D.~J.,  {Robin} A.~C.,  {Reyl{\'e}} C.,  {Schultheis} M.,   {Picaud}
  S.,  2006, \mn@doi [\aap] {10.1051/0004-6361:20053842}, \href
  {https://ui.adsabs.harvard.edu/abs/2006A&A...453..635M} {453, 635}

\bibitem[\protect\citeauthoryear{{Masci} et~al.,}{{Masci}
  et~al.}{2019}]{2019PASP..131a8003M}
{Masci} F.~J.,  et~al., 2019, \mn@doi [\pasp] {10.1088/1538-3873/aae8ac}, \href
  {https://ui.adsabs.harvard.edu/abs/2019PASP..131a8003M} {131, 018003}

\bibitem[\protect\citeauthoryear{{Merc}}{{Merc}}{2025}]{2025Galax..13...49M}
{Merc} J.,  2025, \mn@doi [Galaxies] {10.3390/galaxies13030049}, \href
  {https://ui.adsabs.harvard.edu/abs/2025Galax..13...49M} {13, 49}

\bibitem[\protect\citeauthoryear{{Merc} \& {Miko{\l}ajewska}}{{Merc} \&
  {Miko{\l}ajewska}}{2024}]{2024NatAs...8.1504M}
{Merc} J.,  {Miko{\l}ajewska} J.,  2024, \mn@doi [Nature Astronomy]
  {10.1038/s41550-024-02436-z}, \href
  {https://ui.adsabs.harvard.edu/abs/2024NatAs...8.1504M} {8, 1504}

\bibitem[\protect\citeauthoryear{{Merc}, {G{\'a}lis}  \& {Leedj{\"a}rv}}{{Merc}
  et~al.}{2017}]{2017gacv.workE..60M}
{Merc} J.,  {G{\'a}lis} R.,   {Leedj{\"a}rv} L.,  2017, in The Golden Age of
  Cataclysmic Variables and Related Objects IV. p.~60 (\mn@eprint {arXiv}
  {1806.05935}), \mn@doi{10.22323/1.315.0060}

\bibitem[\protect\citeauthoryear{{Merc}, {G{\'a}lis}  \& {Wolf}}{{Merc}
  et~al.}{2019a}]{2019RNAAS...3...28M}
{Merc} J.,  {G{\'a}lis} R.,   {Wolf} M.,  2019a, \mn@doi [Research Notes of the
  American Astronomical Society] {10.3847/2515-5172/ab0429}, \href
  {https://ui.adsabs.harvard.edu/abs/2019RNAAS...3...28M} {3, 28}

\bibitem[\protect\citeauthoryear{{Merc}, {G{\'a}lis}  \& {Wolf}}{{Merc}
  et~al.}{2019b}]{2019AN....340..598M}
{Merc} J.,  {G{\'a}lis} R.,   {Wolf} M.,  2019b, \mn@doi [Astronomische
  Nachrichten] {10.1002/asna.201913662}, \href
  {https://ui.adsabs.harvard.edu/abs/2019AN....340..598M} {340, 598}

\bibitem[\protect\citeauthoryear{{Merc} et~al.,}{{Merc}
  et~al.}{2020}]{2020A&A...644A..49M}
{Merc} J.,  et~al., 2020, \mn@doi [\aap] {10.1051/0004-6361/202039132}, \href
  {https://ui.adsabs.harvard.edu/abs/2020A&A...644A..49M} {644, A49}

\bibitem[\protect\citeauthoryear{{Merc}, {G{\'a}lis}, {Wolf}, {Velez},
  {Bohlsen}  \& {Barlow}}{{Merc} et~al.}{2022a}]{2022MNRAS.510.1404M}
{Merc} J.,  {G{\'a}lis} R.,  {Wolf} M.,  {Velez} P.,  {Bohlsen} T.,   {Barlow}
  B.~N.,  2022a, \mn@doi [\mnras] {10.1093/mnras/stab3512}, \href
  {https://ui.adsabs.harvard.edu/abs/2022MNRAS.510.1404M} {510, 1404}

\bibitem[\protect\citeauthoryear{{Merc} et~al.,}{{Merc}
  et~al.}{2022b}]{2022ATel15340....1M}
{Merc} J.,  et~al., 2022b, The Astronomer's Telegram, \href
  {https://ui.adsabs.harvard.edu/abs/2022ATel15340....1M} {15340, 1}

\bibitem[\protect\citeauthoryear{{Merc} et~al.,}{{Merc}
  et~al.}{2023}]{2023MNRAS.523..163M}
{Merc} J.,  et~al., 2023, \mn@doi [\mnras] {10.1093/mnras/stad1434}, \href
  {https://ui.adsabs.harvard.edu/abs/2023MNRAS.523..163M} {523, 163}

\bibitem[\protect\citeauthoryear{{Merc}, {Velez}, {Charbonnel}, {Garde}, {Le
  D{\^u}}, {Mulato}, {Petit}  \& {Skowron}}{{Merc}
  et~al.}{2024}]{2024AN....34540017M}
{Merc} J.,  {Velez} P.,  {Charbonnel} S.,  {Garde} O.,  {Le D{\^u}} P.,
  {Mulato} L.,  {Petit} T.,   {Skowron} J.,  2024, \mn@doi [Astronomische
  Nachrichten] {10.1002/asna.20240017}, \href
  {https://ui.adsabs.harvard.edu/abs/2024AN....34540017M} {345, e20240017}

\bibitem[\protect\citeauthoryear{{Merc}, {Mulato}, {Charbonnel}, {Garde}, {Le
  D\^{u}}  \& {Petit}}{{Merc} et~al.}{2025a}]{Merc+Gaia}
{Merc} J.,  {Mulato} L.,  {Charbonnel} S.,  {Garde} O.,  {Le D\^{u}} P.,
  {Petit} T.,  2025a, in preparation

\bibitem[\protect\citeauthoryear{{Merc}, {G{\'a}lis}  \& {Wolf}}{{Merc}
  et~al.}{2025b}]{Merc+NODSV2025}
{Merc} J.,  {G{\'a}lis} R.,   {Wolf} M.,  2025b, submitted to ApJS

\bibitem[\protect\citeauthoryear{{Merc}, {Miko{\l}ajewska}, {Ga{\l}an},
  {I{\l}kiewicz}, {Beck}, {Monard}  \& {Gromadzki}}{{Merc}
  et~al.}{2025c}]{2025arXiv251122988M}
{Merc} J.,  {Miko{\l}ajewska} J.,  {Ga{\l}an} C.,  {I{\l}kiewicz} K.,  {Beck}
  P.~G.,  {Monard} B.,   {Gromadzki} M.,  2025c, arXiv e-prints, \href
  {https://ui.adsabs.harvard.edu/abs/2025arXiv251122988M} {p. arXiv:2511.22988}

\bibitem[\protect\citeauthoryear{{Miko{\l}ajewska}}{{Miko{\l}ajewska}}{2000}]{2000ASPC..199..431M}
{Miko{\l}ajewska} J.,  2000, in {Kastner} J.~H.,  {Soker} N.,   {Rappaport} S.,
   eds,  Astronomical Society of the Pacific Conference Series Vol. 199,
  Asymmetrical Planetary Nebulae II: From Origins to Microstructures. p.~431
  (\mn@eprint {arXiv} {astro-ph/0001012}),
  \mn@doi{10.48550/arXiv.astro-ph/0001012}

\bibitem[\protect\citeauthoryear{{Miko{\l}ajewska}}{{Miko{\l}ajewska}}{2012}]{2012BaltA..21....5M}
{Miko{\l}ajewska} J.,  2012, \mn@doi [Baltic Astronomy]
  {10.1515/astro-2017-0352}, \href
  {https://ui.adsabs.harvard.edu/abs/2012BaltA..21....5M} {21, 5}

\bibitem[\protect\citeauthoryear{{Miko{\l}ajewska}, {Acker}  \&
  {Stenholm}}{{Miko{\l}ajewska} et~al.}{1997}]{1997A&A...327..191M}
{Miko{\l}ajewska} J.,  {Acker} A.,   {Stenholm} B.,  1997, \aap, \href
  {https://ui.adsabs.harvard.edu/abs/1997A&A...327..191M} {327, 191}

\bibitem[\protect\citeauthoryear{{Miko{\l}ajewska}, {Brandi}, {Hack},
  {Whitelock}, {Barba}, {Garcia}  \& {Marang}}{{Miko{\l}ajewska}
  et~al.}{1999}]{1999MNRAS.305..190M}
{Miko{\l}ajewska} J.,  {Brandi} E.,  {Hack} W.,  {Whitelock} P.~A.,  {Barba}
  R.,  {Garcia} L.,   {Marang} F.,  1999, \mn@doi [\mnras]
  {10.1046/j.1365-8711.1999.02449.x}, \href
  {https://ui.adsabs.harvard.edu/abs/1999MNRAS.305..190M} {305, 190}

\bibitem[\protect\citeauthoryear{{Miszalski} \& {Miko{\l}ajewska}}{{Miszalski}
  \& {Miko{\l}ajewska}}{2014}]{2014MNRAS.440.1410M}
{Miszalski} B.,  {Miko{\l}ajewska} J.,  2014, \mn@doi [\mnras]
  {10.1093/mnras/stu292}, \href
  {https://ui.adsabs.harvard.edu/abs/2014MNRAS.440.1410M} {440, 1410}

\bibitem[\protect\citeauthoryear{{Miszalski}, {Miko{\l}ajewska}  \&
  {Udalski}}{{Miszalski} et~al.}{2013}]{2013MNRAS.432.3186M}
{Miszalski} B.,  {Miko{\l}ajewska} J.,   {Udalski} A.,  2013, \mn@doi [\mnras]
  {10.1093/mnras/stt673}, \href
  {https://ui.adsabs.harvard.edu/abs/2013MNRAS.432.3186M} {432, 3186}

\bibitem[\protect\citeauthoryear{{Munari}}{{Munari}}{2019}]{2019arXiv190901389M}
{Munari} U.,  2019, \mn@doi [arXiv e-prints] {10.48550/arXiv.1909.01389}, \href
  {https://ui.adsabs.harvard.edu/abs/2019arXiv190901389M} {p. arXiv:1909.01389}

\bibitem[\protect\citeauthoryear{{Munari} et~al.,}{{Munari}
  et~al.}{2021}]{2021MNRAS.505.6121M}
{Munari} U.,  et~al., 2021, \mn@doi [\mnras] {10.1093/mnras/stab1620}, \href
  {https://ui.adsabs.harvard.edu/abs/2021MNRAS.505.6121M} {505, 6121}

\bibitem[\protect\citeauthoryear{{Murset} \& {Nussbaumer}}{{Murset} \&
  {Nussbaumer}}{1994}]{1994A&A...282..586M}
{Murset} U.,  {Nussbaumer} H.,  1994, \aap, \href
  {https://ui.adsabs.harvard.edu/abs/1994A&A...282..586M} {282, 586}

\bibitem[\protect\citeauthoryear{{O'Donoghue} et~al.,}{{O'Donoghue}
  et~al.}{2006}]{2006MNRAS.372..151O}
{O'Donoghue} D.,  et~al., 2006, \mn@doi [\mnras]
  {10.1111/j.1365-2966.2006.10834.x}, \href
  {https://ui.adsabs.harvard.edu/abs/2006MNRAS.372..151O} {372, 151}

\bibitem[\protect\citeauthoryear{{Parker} et~al.,}{{Parker}
  et~al.}{2005}]{2005MNRAS.362..689P}
{Parker} Q.~A.,  et~al., 2005, \mn@doi [\mnras]
  {10.1111/j.1365-2966.2005.09350.x}, \href
  {https://ui.adsabs.harvard.edu/abs/2005MNRAS.362..689P} {362, 689}

\bibitem[\protect\citeauthoryear{{Rimoldini} et~al.,}{{Rimoldini}
  et~al.}{2023}]{2023A&A...674A..14R}
{Rimoldini} L.,  et~al., 2023, \mn@doi [\aap] {10.1051/0004-6361/202245591},
  \href {https://ui.adsabs.harvard.edu/abs/2023A&A...674A..14R} {674, A14}

\bibitem[\protect\citeauthoryear{{Rodr{\'\i}guez-Flores}, {Corradi}, {Mampaso},
  {Garc{\'\i}a-Alvarez}, {Munari}, {Greimel}, {Rubio-D{\'\i}ez}  \&
  {Santander-Garc{\'\i}a}}{{Rodr{\'\i}guez-Flores}
  et~al.}{2014}]{2014A&A...567A..49R}
{Rodr{\'\i}guez-Flores} E.~R.,  {Corradi} R.~L.~M.,  {Mampaso} A.,
  {Garc{\'\i}a-Alvarez} D.,  {Munari} U.,  {Greimel} R.,  {Rubio-D{\'\i}ez}
  M.~M.,   {Santander-Garc{\'\i}a} M.,  2014, \mn@doi [\aap]
  {10.1051/0004-6361/201323182}, \href
  {https://ui.adsabs.harvard.edu/abs/2014A&A...567A..49R} {567, A49}

\bibitem[\protect\citeauthoryear{{Scargle}}{{Scargle}}{1982}]{1982ApJ...263..835S}
{Scargle} J.~D.,  1982, \mn@doi [\apj] {10.1086/160554}, \href
  {https://ui.adsabs.harvard.edu/abs/1982ApJ...263..835S} {263, 835}

\bibitem[\protect\citeauthoryear{{Schmid}}{{Schmid}}{1989}]{1989A&A...211L..31S}
{Schmid} H.~M.,  1989, \aap, \href
  {https://ui.adsabs.harvard.edu/abs/1989A&A...211L..31S} {211, L31}

\bibitem[\protect\citeauthoryear{{Schwartz}, {Persson}  \& {Hamann}}{{Schwartz}
  et~al.}{1990}]{1990AJ....100..793S}
{Schwartz} R.~D.,  {Persson} S.~E.,   {Hamann} F.~W.,  1990, \mn@doi [\aj]
  {10.1086/115561}, \href
  {https://ui.adsabs.harvard.edu/abs/1990AJ....100..793S} {100, 793}

\bibitem[\protect\citeauthoryear{{Shappee} et~al.,}{{Shappee}
  et~al.}{2014}]{2014ApJ...788...48S}
{Shappee} B.~J.,  et~al., 2014, \mn@doi [\apj] {10.1088/0004-637X/788/1/48},
  \href {https://ui.adsabs.harvard.edu/abs/2014ApJ...788...48S} {788, 48}

\bibitem[\protect\citeauthoryear{{Shingles} et~al.,}{{Shingles}
  et~al.}{2021}]{2021TNSAN...7....1S}
{Shingles} L.,  et~al., 2021, Transient Name Server AstroNote, \href
  {https://ui.adsabs.harvard.edu/abs/2021TNSAN...7....1S} {7, 1}

\bibitem[\protect\citeauthoryear{{Skrutskie} et~al.,}{{Skrutskie}
  et~al.}{2006}]{2006AJ....131.1163S}
{Skrutskie} M.~F.,  et~al., 2006, \mn@doi [\aj] {10.1086/498708}, \href
  {https://ui.adsabs.harvard.edu/abs/2006AJ....131.1163S} {131, 1163}

\bibitem[\protect\citeauthoryear{{Smith} et~al.,}{{Smith}
  et~al.}{2020}]{2020PASP..132h5002S}
{Smith} K.~W.,  et~al., 2020, \mn@doi [\pasp] {10.1088/1538-3873/ab936e}, \href
  {https://ui.adsabs.harvard.edu/abs/2020PASP..132h5002S} {132, 085002}

\bibitem[\protect\citeauthoryear{{Sokoloski} et~al.,}{{Sokoloski}
  et~al.}{2006}]{2006ApJ...636.1002S}
{Sokoloski} J.~L.,  et~al., 2006, \mn@doi [\apj] {10.1086/498206}, \href
  {https://ui.adsabs.harvard.edu/abs/2006ApJ...636.1002S} {636, 1002}

\bibitem[\protect\citeauthoryear{{Terzan}, {Bernard}  \& {Guibert}}{{Terzan}
  et~al.}{1997}]{1997A&AS..123..507T}
{Terzan} A.,  {Bernard} A.,   {Guibert} J.,  1997, \mn@doi [\aaps]
  {10.1051/aas:1997351}, \href
  {https://ui.adsabs.harvard.edu/abs/1997A&AS..123..507T} {123, 507}

\bibitem[\protect\citeauthoryear{{Tonry} et~al.,}{{Tonry}
  et~al.}{2018}]{2018PASP..130f4505T}
{Tonry} J.~L.,  et~al., 2018, \mn@doi [\pasp] {10.1088/1538-3873/aabadf}, \href
  {https://ui.adsabs.harvard.edu/abs/2018PASP..130f4505T} {130, 064505}

\bibitem[\protect\citeauthoryear{{Udalski}, {Szymanski}, {Soszynski}  \&
  {Poleski}}{{Udalski} et~al.}{2008}]{2008AcA....58...69U}
{Udalski} A.,  {Szymanski} M.~K.,  {Soszynski} I.,   {Poleski} R.,  2008,
  \mn@doi [\actaa] {10.48550/arXiv.0807.3884}, \href
  {https://ui.adsabs.harvard.edu/abs/2008AcA....58...69U} {58, 69}

\bibitem[\protect\citeauthoryear{{Udalski}, {Szyma{\'n}ski}  \&
  {Szyma{\'n}ski}}{{Udalski} et~al.}{2015}]{2015AcA....65....1U}
{Udalski} A.,  {Szyma{\'n}ski} M.~K.,   {Szyma{\'n}ski} G.,  2015, \mn@doi
  [\actaa] {10.48550/arXiv.1504.05966}, \href
  {https://ui.adsabs.harvard.edu/abs/2015AcA....65....1U} {65, 1}

\bibitem[\protect\citeauthoryear{{VanderPlas} \& {Ivezi{\'c}}}{{VanderPlas} \&
  {Ivezi{\'c}}}{2015}]{2015ApJ...812...18V}
{VanderPlas} J.~T.,  {Ivezi{\'c}} {\v{Z}}.,  2015, \mn@doi [\apj]
  {10.1088/0004-637X/812/1/18}, \href
  {https://ui.adsabs.harvard.edu/abs/2015ApJ...812...18V} {812, 18}

\bibitem[\protect\citeauthoryear{{Vioque}, {Oudmaijer}, {Schreiner},
  {Mendigut{\'\i}a}, {Baines}, {Mowlavi}  \&
  {P{\'e}rez-Mart{\'\i}nez}}{{Vioque} et~al.}{2020}]{2020A&A...638A..21V}
{Vioque} M.,  {Oudmaijer} R.~D.,  {Schreiner} M.,  {Mendigut{\'\i}a} I.,
  {Baines} D.,  {Mowlavi} N.,   {P{\'e}rez-Mart{\'\i}nez} R.,  2020, \mn@doi
  [\aap] {10.1051/0004-6361/202037731}, \href
  {https://ui.adsabs.harvard.edu/abs/2020A&A...638A..21V} {638, A21}

\bibitem[\protect\citeauthoryear{{Wray}}{{Wray}}{1966}]{1966PhDT.........3W}
{Wray} J.~D.,  1966, PhD thesis, Northwestern University

\bibitem[\protect\citeauthoryear{{Wright} et~al.,}{{Wright}
  et~al.}{2010}]{2010AJ....140.1868W}
{Wright} E.~L.,  et~al., 2010, \mn@doi [\aj] {10.1088/0004-6256/140/6/1868},
  \href {https://ui.adsabs.harvard.edu/abs/2010AJ....140.1868W} {140, 1868}

\bibitem[\protect\citeauthoryear{{Xu}, {Shao}  \& {Li}}{{Xu}
  et~al.}{2024}]{2024ApJ...962..126X}
{Xu} X.-j.,  {Shao} Y.,   {Li} X.-D.,  2024, \mn@doi [\apj]
  {10.3847/1538-4357/ad20ec}, \href
  {https://ui.adsabs.harvard.edu/abs/2024ApJ...962..126X} {962, 126}

\bibitem[\protect\citeauthoryear{{Zhao}, {Guo}, {Lv}, {Li}, {Zhu}  \&
  {Shi}}{{Zhao} et~al.}{2025}]{2025arXiv250720206Z}
{Zhao} Y.,  {Guo} S.,  {Lv} G.,  {Li} J.,  {Zhu} C.,   {Shi} J.,  2025, \mn@doi
  [arXiv e-prints] {10.48550/arXiv.2507.20206}, \href
  {https://ui.adsabs.harvard.edu/abs/2025arXiv250720206Z} {p. arXiv:2507.20206}

\bibitem[\protect\citeauthoryear{{van Belle} et~al.,}{{van Belle}
  et~al.}{1999}]{1999AJ....117..521V}
{van Belle} G.~T.,  et~al., 1999, \mn@doi [\aj] {10.1086/300677}, \href
  {https://ui.adsabs.harvard.edu/abs/1999AJ....117..521V} {117, 521}

\makeatother
\end{thebibliography}




\bsp	
\label{lastpage}

\appendix

\section{Symbiotic stars from Paper~I}

The list of bona-fide and possible symbiotic stars from \citetalias{2014MNRAS.440.1410M}, together with their SIMBAD and \textit{Gaia} DR3 identifiers, galactic coordinates, parallaxes, distances and \textit{Gaia} magnitudes is in Table \ref{tab:basic_data_paperI}. Their SALT/RSS spectra are shown in figs. 1-4 of the aforementioned paper. The infrared 2MASS magnitudes, together with their infrared types, spectral types of the giants, and their effective temperatures are in Table \ref{tab:2mass_paperI}. The fluxes of emission lines and the temperatures of the hot components are listed in Table \ref{tab:nebular_paperI}. Information on their variability is in \ref{tab:variability_paperI}.

\begin{table*}
	\centering
	\caption{The list of the bona-fide and possible symbiotic stars from \citetalias{2014MNRAS.440.1410M}. The columns are the same as in Table \ref{tab:basic_data}.}
\label{tab:basic_data_paperI}
\begin{tabular}{lllrrrrr}
\hline

Name (2MASS~J) & SIMBAD name & \textit{Gaia} DR3 & OGLE IV & $\ell$  & $b$  & $\varpi$  & d  \\
 &  &  &  & (\degr) & (\degr) & (mas) & (kpc) \\
\hline
\textit{Bona-fide} &  &  &  &  &  &  &  \\
14031865-5809349 & WRAY 15-1167 & 5871060800085618816 & GD1247.27.6 & 312.3148 & 3.3967 & 0.09$\pm$0.02 & 7.0 \\
15431767-5857221 & 2MASS~J154.. & 5834224583611758592 & GD1181.19.24296 & 323.5413 & -3.1423 & 0.00$\pm$0.02 & 12.8 \\
16003761-4835228 & 2MASS~J160.. & 5984320908710818944 & GD1152.19.59 & 332.0679 & 3.2823 & 0.08$\pm$0.03 & 6.1 \\
16422739-4133105 & IRAS 16389-4127 & 5968218384802364032 & GD1102.09.592 & 342.2640 & 3.0318 & 0.00$\pm$0.07 & 7.8 \\
17050868-4849122 & 2MASS~J170.. & 5938619188187996800 & GD1916.27.23710 & 339.1468 & -4.6493 & 0.07$\pm$0.03 & 6.6 \\
17334728-2719266 & Terz V 2513 & 4061345440488592896$^*$ & BLG611.15.9091$^*$ & 359.9792 & 3.0664 & -0.29$\pm$0.13 & 6.7 \\
17391715-3546593 & PN K 5-8 & 4041216852950479616 & BLG610.21.103715 & 353.4730 & -2.4679 & -0.17$\pm$0.14 & 10.1 \\
17422035-2401162 & OGLE BLG-LPV-19199 & 4068379669255286912 & BLG626.18.75294 & 3.8061 & 3.1974 & -0.04$\pm$0.05 & 8.3 \\
17463311-2419558 & 2MASS~J174.. & 4068133481744764032 & BLG633.28.25 & 4.0423 & 2.2155 & 0.14$\pm$0.05 & 5.7 \\
18131474-1007218 & 2MASS~J181.. & 4157296551738802432 & - & 19.5540 & 3.7375 & 0.05$\pm$0.04 & 6.5 \\
18272892-1555547 & 2MASS~J182.. & 4097140453884745088 & BLG576.15.38 & 16.0601 & -2.0558 & 0.19$\pm$0.08 & 4.1 \\
18300636-1940315 & 2MASS~J183.. & 4092813080982916736 & BLG579.11.85 & 13.0225 & -4.3378 & -0.08$\pm$0.04 & 9.7 \\
 &  &  &  &  &  &  &  \\
\textit{Possible} &  &  &  &  &  &  &  \\
16503229-4742288 & IRAS 16468-4737 & 5939392802025441152 & GD1100.27.1204 & 338.5095 & -2.0533 & 0.14$\pm$0.18 & 4.0 \\
17145509-3933117 & IRAS 17114-3929 & 5972229544962050432 & BLG990.18.2562 & 347.6561 & -0.5638 & 0.18$\pm$0.10 & 3.9 \\
17460199-3303085 & 2MASS~J174.. & 4053970672432264192 & - & 356.5300 & -2.2173 & 0.04$\pm$0.03 & 7.7 \\
\hline
\end{tabular}

 \medskip
    \begin{minipage}{\linewidth}
    \textbf{Notes.} $^{*}$Although another \textit{Gaia} source (and OGLE source) lies closer to the 2MASS position, the analysis of \citet{2025arXiv251122988M} confirms that the source ID listed here corresponds to the symbiotic star. See the text for further details.
    \end{minipage}
\end{table*}

 \begin{table*}
 \setlength{\tabcolsep}{5.1pt}
	\centering
	\caption{\textit{Gaia} DR3 magnitudes, 2MASS near-IR magnitudes, extinction, IR types, giant spectral types and effective temperatures of bona-fide and possible symbiotic stars from \citetalias{2014MNRAS.440.1410M}. The columns are the same as in Table \ref{tab:2mass}.  }
	\label{tab:2mass_paperI}
\begin{tabular}{lrrr|rrrrccr}
\hline

Name & $G_{\rm BP}$ & $G$ & $G_{\rm RP}$ & $J$ & $H$ & $K_s$ & E$_{\rm(B-V)}$ & IR & Giant & T$_{\rm eff}$ \\
 & (mag) & (mag) & (mag) & (mag) & (mag) & (mag) & (mag) & type & SpT & (K) \\ \hline
\textit{Bona-fide} &  &  &  &  &  &  &  &  &  &  \\
14031.. & 15.6 & 13.9 & 12.6 & 10.42 & 9.32 & 8.94 & 0.6 & S & M4 & 3476 \\
15431.. & 16.4 & 14.7 & 13.5 & 11.22 & 10.13 & 9.72 & 0.6 & S & M2.5 & 3641 \\
16003.. & 16.8 & 14.7 & 13.3 & 10.8 & 9.47 & 8.89 & 1.3 & S & C-N5 C24.5 & - \\
16422.. & 17.1 & 16.4 & 15.4 & 10.33 & 8.96 & 7.66 & 1.2 & D & - & - \\
17050.. & 15.9 & 13.9 & 12.5 & 10.04 & 8.95 & 8.57 & 0.6 & S & M4 & 3476 \\
17334.. & 16.6 & 14.9 & 12.5 & 9.32 & 7.91 & 7.26 & 1.6 & S & M2 & 3695 \\
17391.. & 18.2 & 16.6 & 15.1 & 9.81 & 8.44 & 7.63 & 1.2 & S & M1.5 & 3750 \\
17422.. & 17.2 & 15.0 & 13.6 & 10.2 & 8.99 & 8.43 & 1.0 & S & M2: & 3695 \\
17463.. & 17.6 & 15.1 & 13.6 & 9.86 & 8.51 & 7.91 & 1.3 & S & M4: & 3476 \\
18131.. & 17.3 & 15.0 & 13.7 & 10.94 & 9.64 & 9.18 & 1.3 & S & M0 & 3914 \\
18272.. & 16.9 & 14.3 & 12.8 & 9.16 & 7.8 & 7.18 & 1.1 & S & M1 & 3804 \\
18300.. & 16.7 & 15.0 & 13.7 & 11.07 & 9.96 & 9.55 & 0.8 & S & M3.5 & 3531 \\
 &  &  &  &  &  &  &  &  &  &  \\
\textit{Possible} &  &  &  &  &  &  &  &  &  &  \\
16503.. & 19.4 & 18.5 & 17.4 & 13.99 & 11.55 & 9.48 & 1.5 & D? & - & - \\
17145.. & 17.8 & 17.3 & 16.1 & 11.92 & 9.62 & 7.87 & 0.9 & D? & - & - \\
17460.. & 16.4 & 14.7 & 13.3 & 9.86 & 8.7 & 8.17 & 1.4 & S? & K5-M0 & 3969\\
\hline
\end{tabular}
\end{table*}

\begin{table*}
	\centering
     \setlength{\tabcolsep}{5.1pt}
	\caption{Fluxes of emission lines in the spectra of new and possible symbiotic stars from \citetalias{2014MNRAS.440.1410M}. The columns are the same as in Table \ref{tab:nebular}.}
	\label{tab:nebular_paperI}
\begin{tabular}{llllllllllllllll}
\hline

Name & H$\gamma$ & {[}\ion{O}{iii}{]}  & \ion{He}{ii}  & H$\beta$ & {[}\ion{O}{iii}{]}  & {[}\ion{Fe}{vii}{]} & \ion{He}{i}  & {[}\ion{Fe}{vii}{]} & H$\alpha$ & \ion{He}{i}   & \ion{O}{vi}  & \ion{He}{i}  & Ion & T$^{a}_{\rm h}$ & T$^{b}_{\rm h}$ \\
 & {\scriptsize 4340 \AA} & {\scriptsize 4363 \AA}& {\scriptsize 4686 \AA} & {\scriptsize 4861 \AA}& {\scriptsize 5007 \AA} & {\scriptsize 5721 \AA} & {\scriptsize 5876 \AA}& {\scriptsize 6086 \AA}& {\scriptsize 6563 \AA}& {\scriptsize 6678 \AA}& {\scriptsize 6825 \AA}& {\scriptsize 7065 \AA}&  & (10$^3$ K) & (10$^3$ K) \\ \hline
\textit{Bona-fide} &  &  &  &  &  &  &  \\
14031.. & - & 7 & 40 & 100 & 7 & 0 & 21 & 1 & 539 & 18 & 0 & 21 & Fe$^{+6}$: & 99 & 141 \\
15431.. & 30 & 0 & 83 & 100 & 0 & 11 & 14 & 16 & 780 & 28 & 77 & 10 & O$^{+5}$ & 114 & 180 \\
16003.. & 22 & 0 & 56 & 100 & 0 & 9 & 16 & 0 & 928 & 13 & 80 & 20 & O$^{+5}$ & 114 & 157 \\
16422.. & 39 & 29 & 73 & 100 & 144 & 27 & 9 & 48 & 477 & 3 & 20 & 8 & O$^{+5}$ & 114 & 172 \\
17050.. & - & 3 & 28 & 100 & 3 & 2 & 20 & 4 & 562 & 21 & 5 & 20 & O$^{+5}$ & 114 & 126 \\
17334.. & - & 53 & 12 & 100 & 119 & 0 & 34 & 0 & 638 & 12 & 0 & 31 & He$^{+2}$ & 54 & 100 \\
17391.. & - & 7 & 62 & 100 & 16 & 24 & 21 & 41 & 945 & 9 & 15 & 20 & O$^{+5}$ & 114 & 163 \\
17422.. & - & 0 & 51 & 100 & 17 & 10 & 24 & 15 & 761 & 16 & 6 & 20 & O$^{+5}$ & 114 & 152 \\
17463.. & - & 4 & 73 & 100 & 58 & 28 & 48 & 52 & 971 & 19 & 0 & 51 & Fe$^{+6}$ & 99 & 172 \\
18131.. & - & 6 & 38 & 100 & 4 & 1 & 30 & 0 & 606 & 40 & 12 & 30 & O$^{+5}$ & 114 & 139 \\
18272.. & - & 0 & 43 & 100 & 0 & 4 & 39 & 7 & 1170 & 36 & 27 & 44 & O$^{+5}$ & 114 & 144 \\
18300.. & - & 9 & 123 & 100 & 0 & 2 & 27 & 7 & 926 & 9 & 86 & 24 & O$^{+5}$ & 114 & 208 \\
 &  &  &  &  &  &  &  \\
\textit{Possible} &  &  &  &  &  &  &  \\
16503.. & - & 107 & 0 & 100 & 759 & 0 & 58 & 0 & 1301 & 18 & 0 & 67 & O$^{+2}$ & 14 & - \\
17145.. & - & 52 & 0 & 100 & 416 & 0 & 41 & 0 & 757 & 14 & 0 & 50 & O$^{+2}$ & 14 & - \\
17460.. & - & 4 & 0 & 100 & 19 & 0 & 31 & 0 & 840 & 16 & 0 & 18 & O$^{+2}$ & 14 & - \\
\hline
\end{tabular}

 \medskip
    \begin{minipage}{\linewidth}
    \textbf{Notes.} $^{a}$Temperature of the hot component estimated from the maximum ionization potential observed in the optical spectrum.
$^{b}$Temperature of the hot component estimated using the method of \citet{1981ASIC...69..517I}. For further details, see the main text.
    \end{minipage}
\end{table*}

\begin{table*}
	\centering
	\caption{Variability of bona-fide and possible symbiotic stars from \citetalias{2014MNRAS.440.1410M}.}
	\label{tab:variability_paperI}

\begin{tabular}{lrrrcc}
\hline
Name & P$_{\rm orb}$ & P$_{\rm pul}$ & P$_{\rm other}$& Outbursts & Eclipses \\
 & (days) & (days) & (days)&  &  \\ \hline
\textit{Bona-fide} &  &  &  &   \\
14031.. & 685$^a$ &  & 563: & Yes \\
15431.. & 592: & 61 &  \\
16003.. & 388 & 51: &  \\
16422.. &  & 369 &  \\
17050.. & 824:$^b$ &  &   \\
17334.. &  & 161$^c$ &  \\
17391.. &  & 153 & 3200: \\
17422.. & 1056 & 90 &  \\
17463.. & 968 & 52 &  \\
18131.. & 446 &  &  \\
18272.. & 1261 &  &  \\
18300.. & 1353 & 78 & 506\\
 &  &  &  &  \\
 \textit{Possible} &  &  &  &   \\
16503.. &  &  & 35$^d$ \\
17145.. &  &  &  \\
17460.. & 729 &  &\\
\hline
\end{tabular}

 \medskip
    \begin{minipage}{\linewidth}
    \textbf{Notes.} $^{a}$Orbital period inferred from \textit{Gaia} $G$-band data available through the \textit{Gaia} Science Alerts website (see Sect.~\ref{sec:individual} for details). 
    $^{b}$Comparison of the $V$- and $I$-band data suggests the presence of ellipsoidal variability, typically interpreted as due to a tidally distorted giant. At longer wavelengths, where the giant dominates the light (unlike shorter wavelengths dominated by the nebula and/or hot companion), two minima and maxima per orbital period are expected \citep[see, e.g., Fig.~6.2 in][]{2019arXiv190901389M}. Consequently, we adopt twice the dominant period as the orbital period.
$^{c}$Value adopted from the analysis of \citet{2025arXiv251122988M}, see text for further discussion.
$^{d}$If the object is indeed a D-type symbiotic system, the 35-day period detected in the OGLE-IV $I$-band data would be unusually short for the pulsation period of the cool Mira component.
    \end{minipage}
    
\end{table*}

\section{Photometry}
In this appendix, we present light curves and periodograms of all studied symbiotic stars and candidates. Figure \ref{fig:photometry} shows data for the newly identified symbiotic stars from this work, while Fig.~\ref{fig:photometry5} presents data for the new candidates. Symbiotic stars and candidates from \citetalias{2014MNRAS.440.1410M} are shown in Fig.~\ref{fig:photometry8} and Fig.~\ref{fig:photometry10}, respectively.

\clearpage

 \begin{figure*}
\centering
 \includegraphics[width=0.469\textwidth]{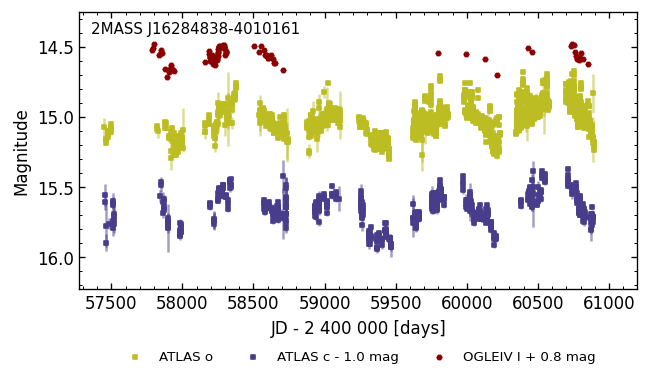}
 \includegraphics[width=0.40\textwidth]{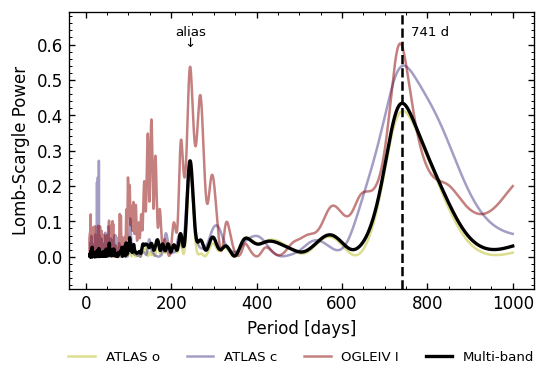}
  \includegraphics[width=0.469\textwidth]{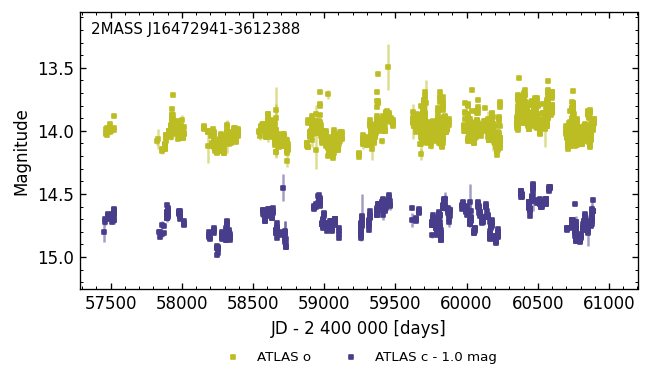}
 \includegraphics[width=0.40\textwidth]{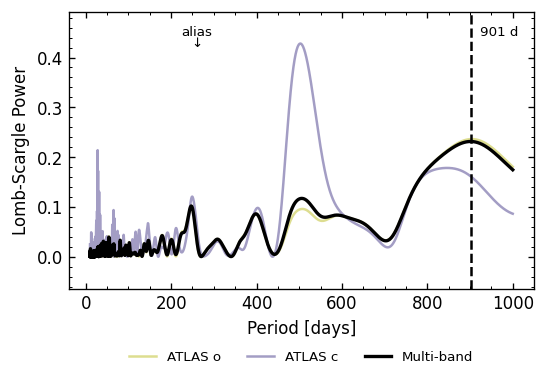}
 \includegraphics[width=0.469\textwidth]{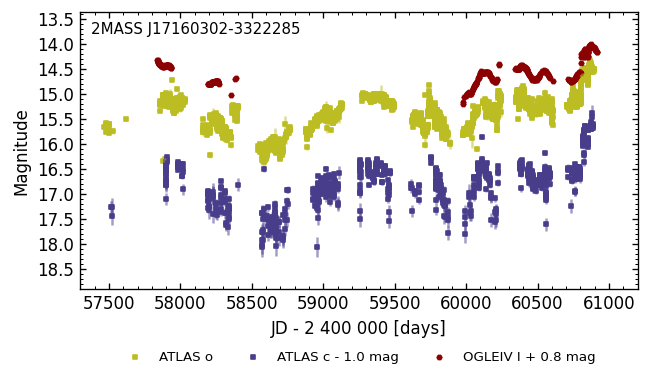}
 \includegraphics[width=0.40\textwidth]{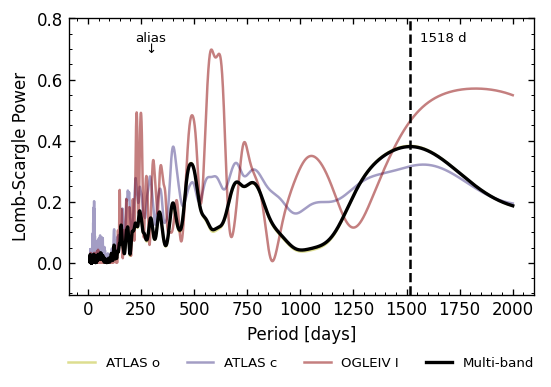}
  \includegraphics[width=0.469\textwidth]{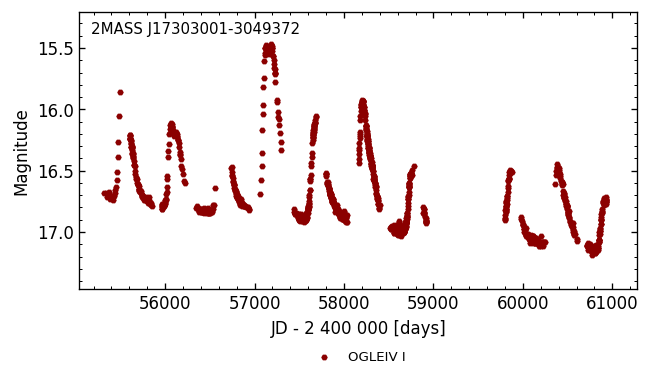}
 \includegraphics[width=0.40\textwidth]{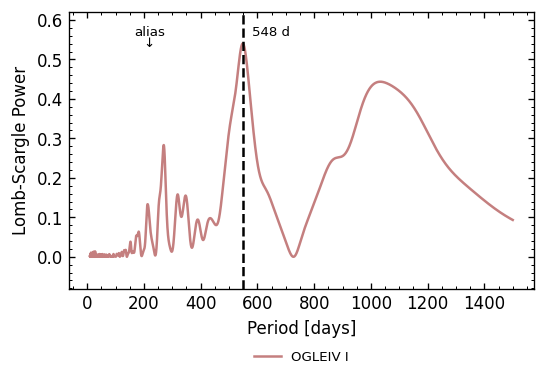}
 \caption{Light curves and Lomb–Scargle periodograms of the new symbiotic stars from this work. For visual clarity, some datasets are shifted by a constant value, as indicated in the legend of each panel. In the periodogram panels, the most prominent period, typically obtained from the multi-band analysis, is marked by a vertical dashed line, with its yearly aliases also indicated. }
 \label{fig:photometry}
 \end{figure*}

\begin{figure*}
\centering

 \includegraphics[width=0.469\textwidth]{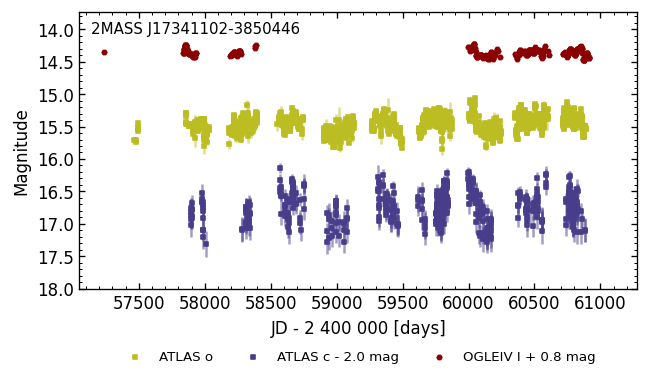}
 \includegraphics[width=0.40\textwidth]{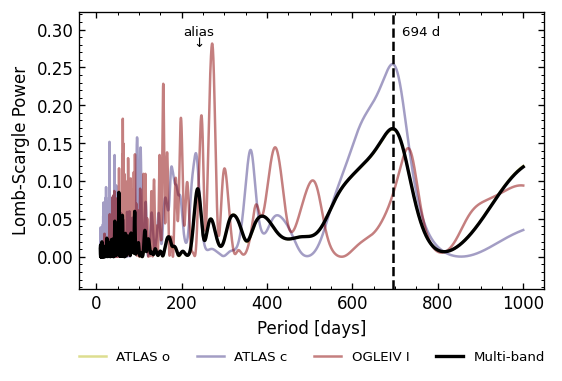}
 
\includegraphics[width=0.469\textwidth]{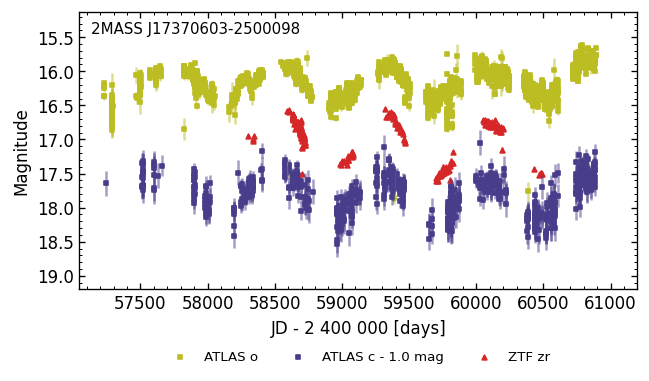}
 \includegraphics[width=0.40\textwidth]{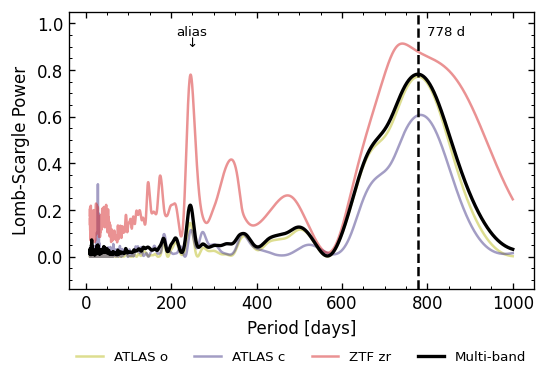}
 
 \includegraphics[width=0.469\textwidth]{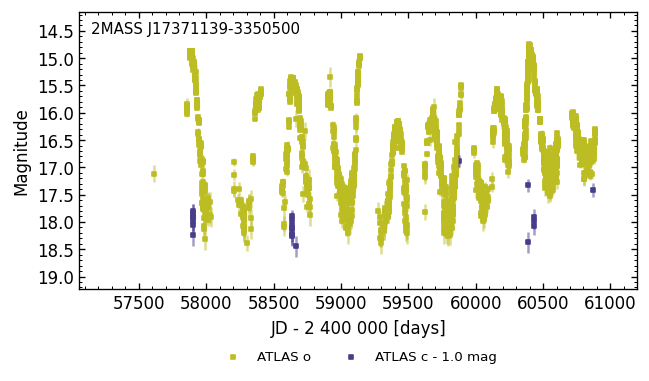}
 \includegraphics[width=0.40\textwidth]{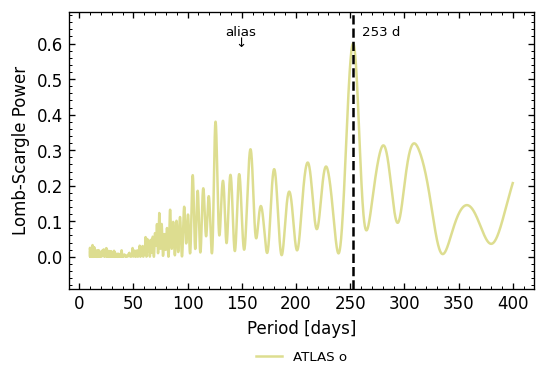}

  \includegraphics[width=0.469\textwidth]{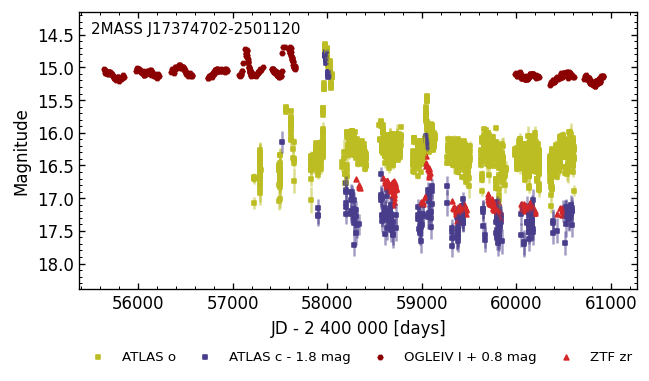}
 \includegraphics[width=0.40\textwidth]{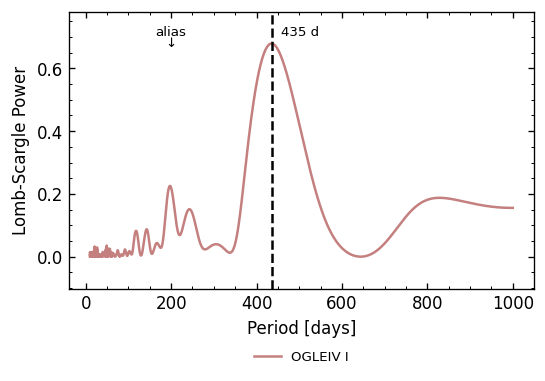}
 
\contcaption{Light curves and Lomb–Scargle periodograms of the new symbiotic stars from this work.}
 \label{fig:photometry2}
 \end{figure*}

 \begin{figure*}
\centering

 \includegraphics[width=0.469\textwidth]{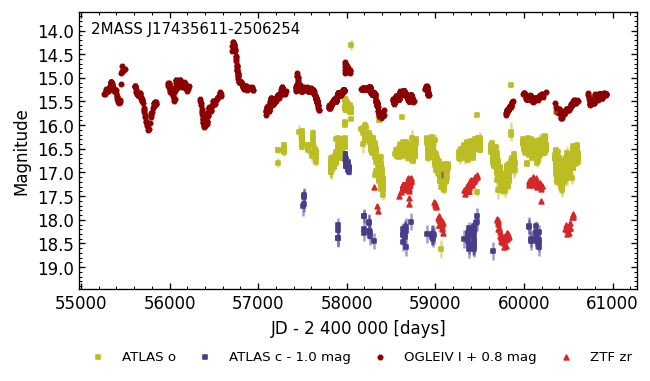}
 \includegraphics[width=0.40\textwidth]{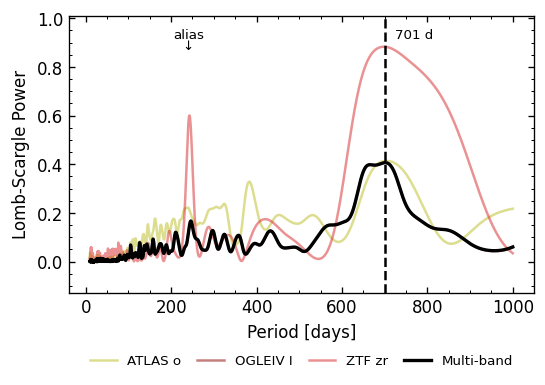}
  \includegraphics[width=0.469\textwidth]{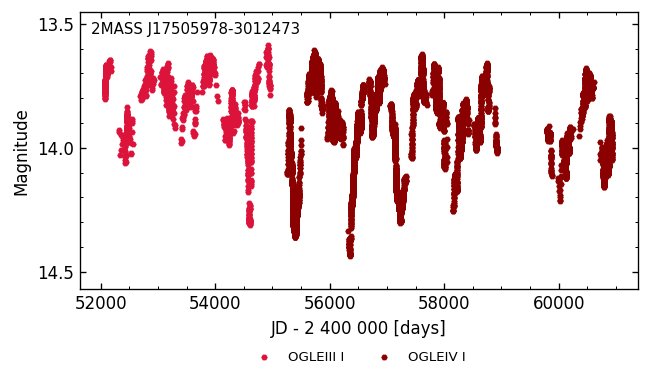}
 \includegraphics[width=0.40\textwidth]{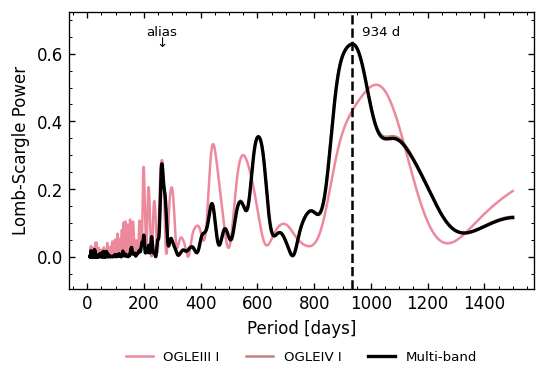}
  \includegraphics[width=0.469\textwidth]{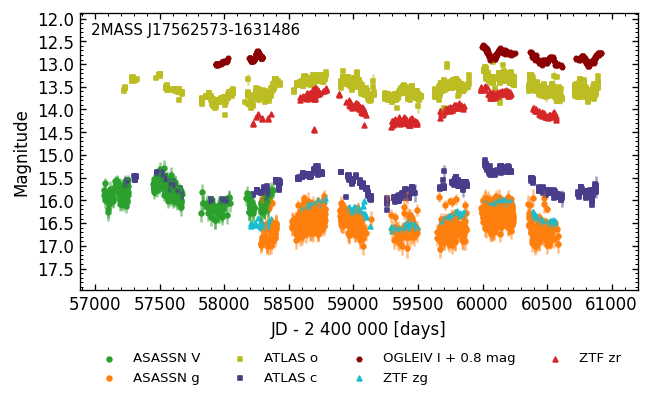}
 \includegraphics[width=0.40\textwidth]{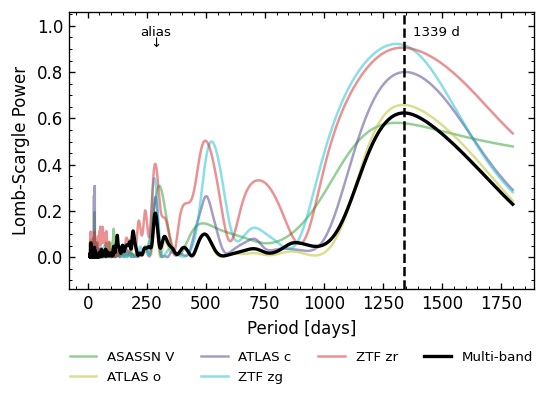}

  \includegraphics[width=0.469\textwidth]{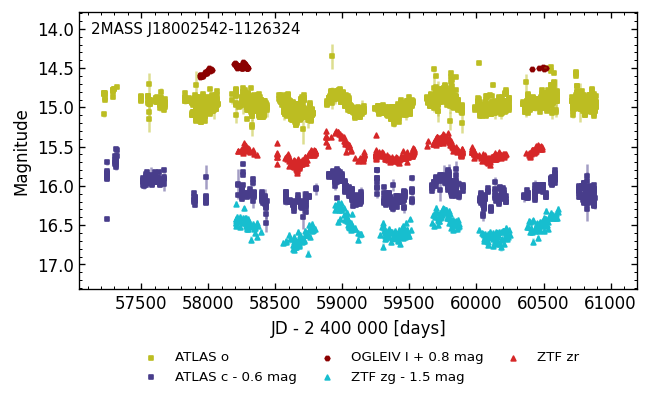}
 \includegraphics[width=0.40\textwidth]{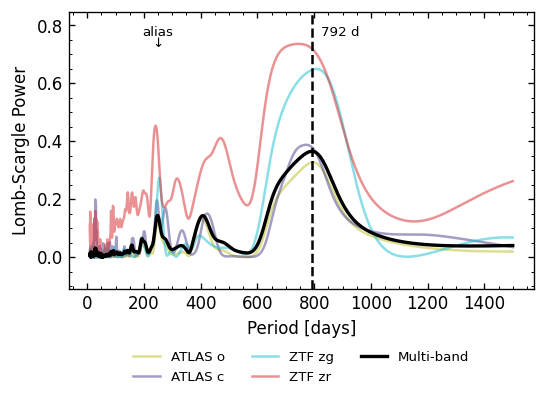}

\contcaption{Light curves and Lomb–Scargle periodograms of the new symbiotic stars from this work.}
 \label{fig:photometry3}
 \end{figure*}

  \begin{figure*}
\centering
 \includegraphics[width=0.469\textwidth]{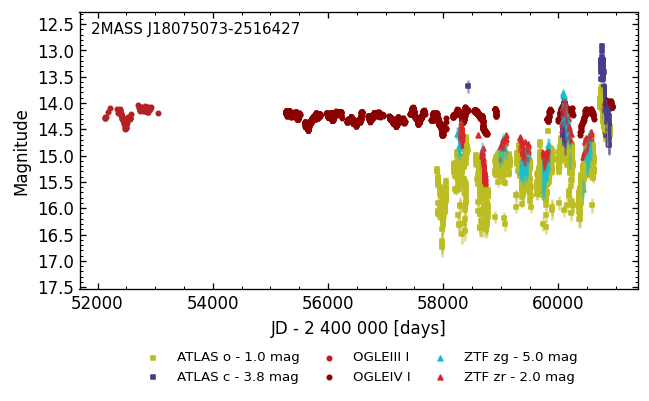}
 \includegraphics[width=0.40\textwidth]{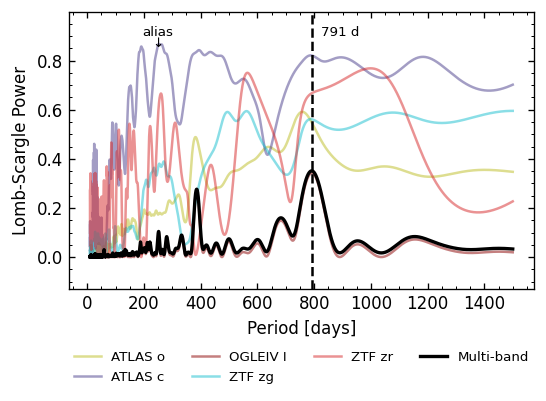}

  \includegraphics[width=0.469\textwidth]{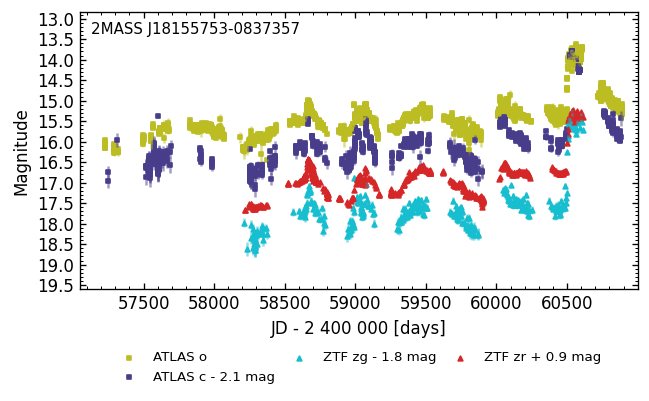}
 \includegraphics[width=0.40\textwidth]{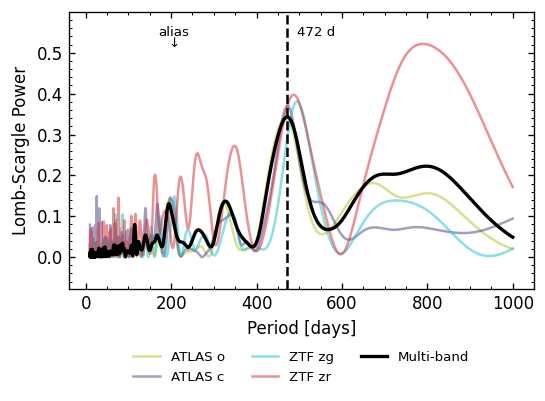}

\contcaption{Light curves and Lomb–Scargle periodograms of the new symbiotic stars from this work.}
 \label{fig:photometry4}
 \end{figure*}

  \begin{figure*}
\centering
 \includegraphics[width=0.469\textwidth]{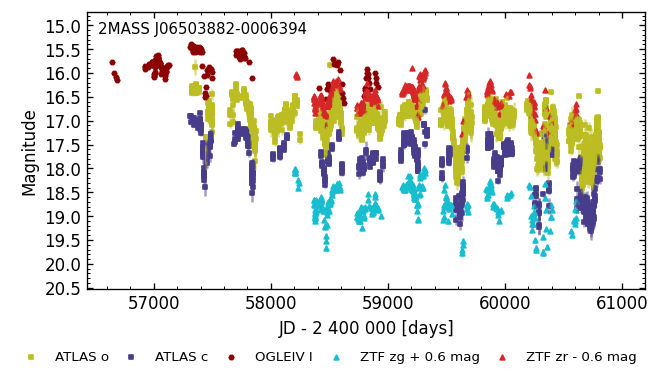}
 \includegraphics[width=0.40\textwidth]{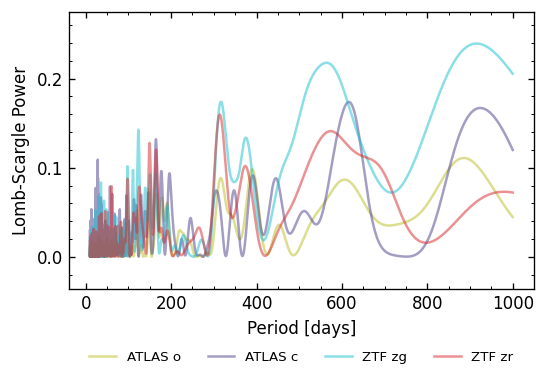}
  \includegraphics[width=0.469\textwidth]{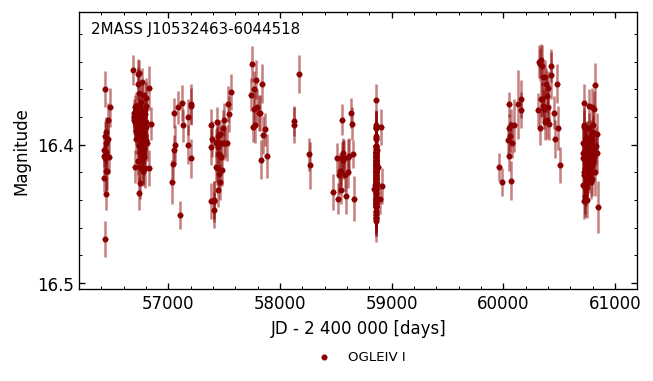}
 \includegraphics[width=0.40\textwidth]{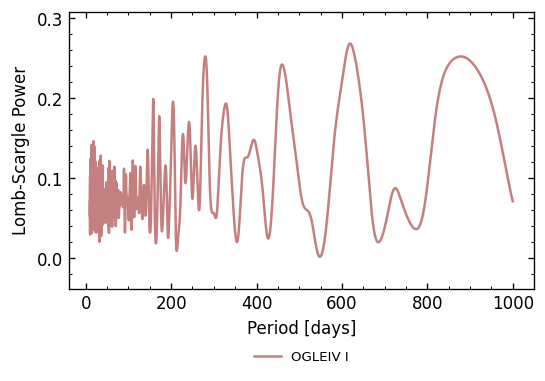}

 \caption{Light curves and Lomb–Scargle periodograms of the possible symbiotic stars from this work.}
 \label{fig:photometry5}
 \end{figure*}

   \begin{figure*}
\centering

  \includegraphics[width=0.469\textwidth]{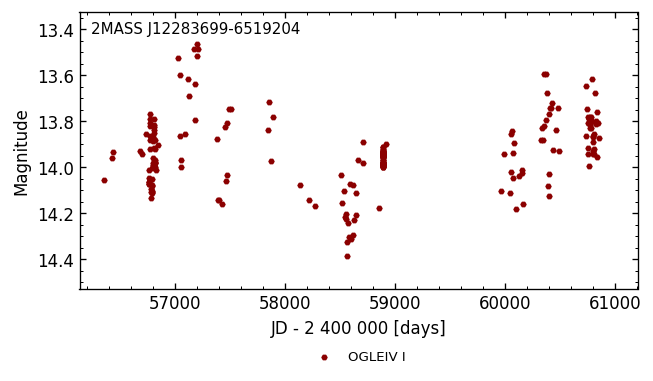}
 \includegraphics[width=0.40\textwidth]{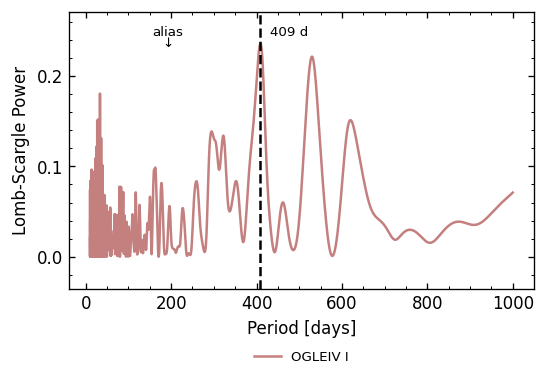}
 
 \includegraphics[width=0.469\textwidth]{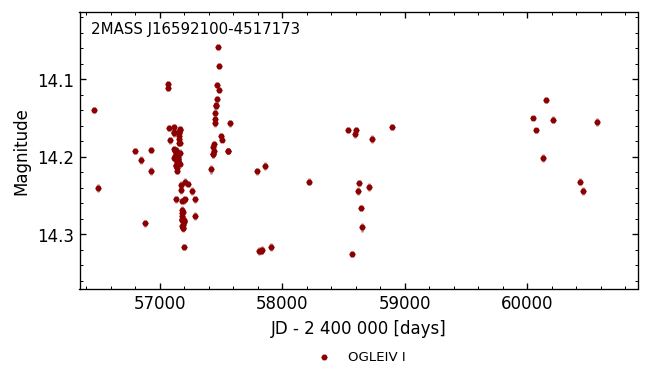}
 \includegraphics[width=0.40\textwidth]{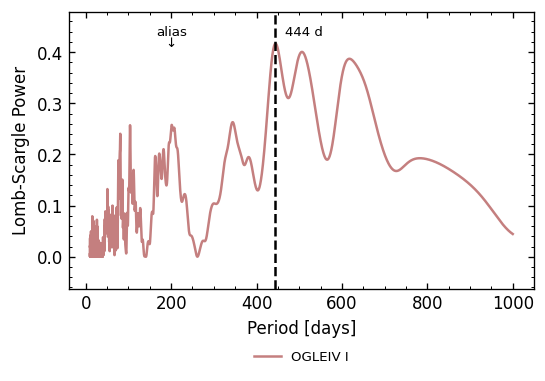}

  \includegraphics[width=0.469\textwidth]{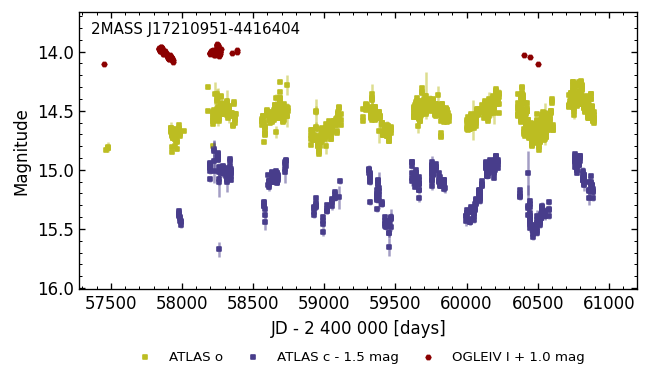}
 \includegraphics[width=0.40\textwidth]{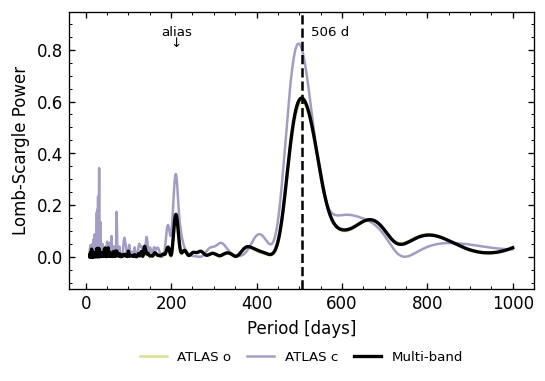}

  \includegraphics[width=0.469\textwidth]{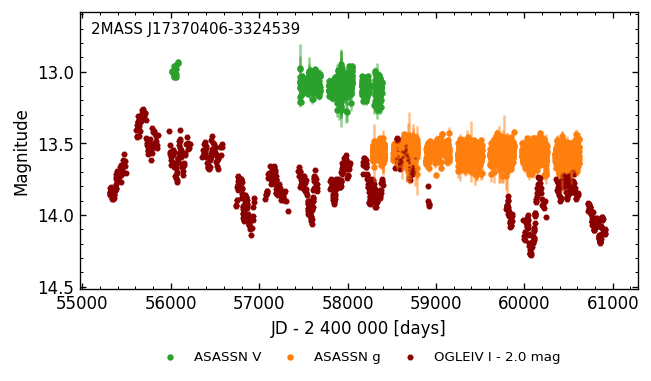}
 \includegraphics[width=0.40\textwidth]{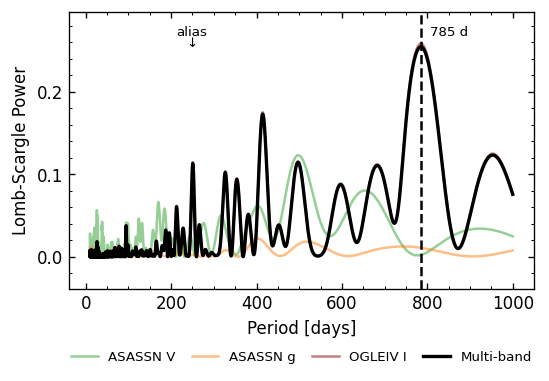}

 \contcaption{Light curves and Lomb–Scargle periodograms of the possible symbiotic stars from this work.}
 \label{fig:photometry6}
 \end{figure*}

    \begin{figure*}
\centering
 \includegraphics[width=0.469\textwidth]{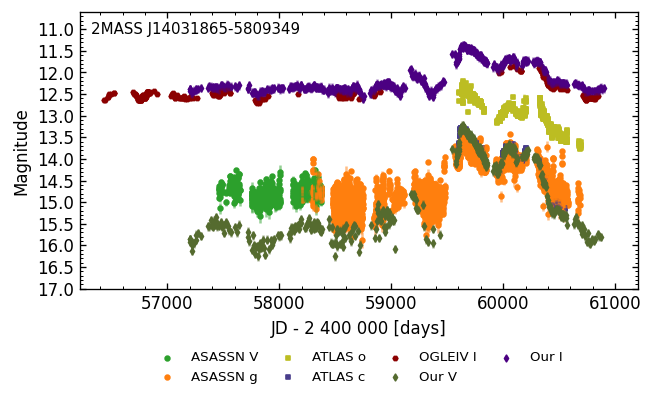}
 \includegraphics[width=0.40\textwidth]{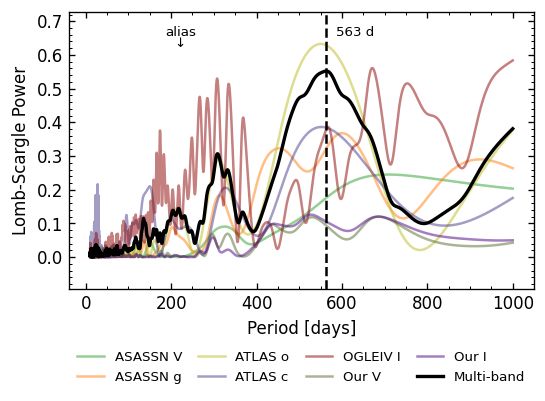}
  \includegraphics[width=0.469\textwidth]{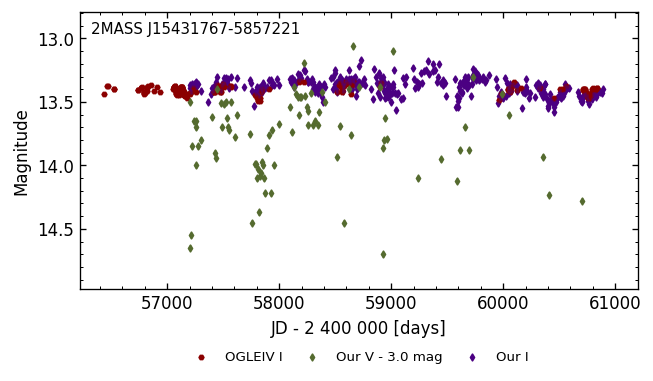}
 \includegraphics[width=0.40\textwidth]{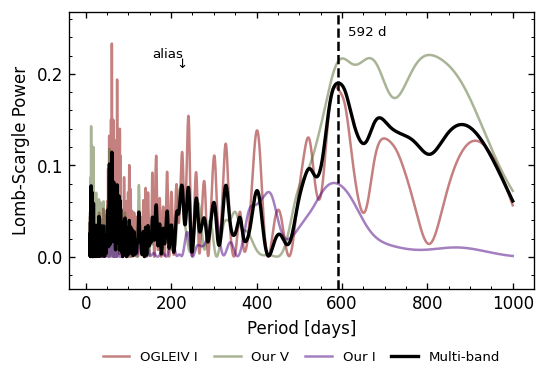}

 \includegraphics[width=0.469\textwidth]{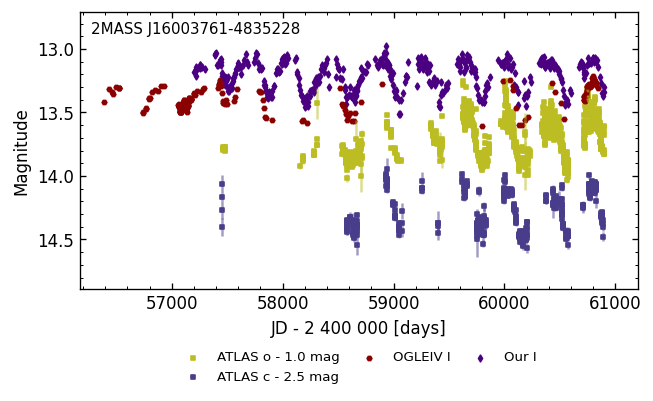}
 \includegraphics[width=0.40\textwidth]{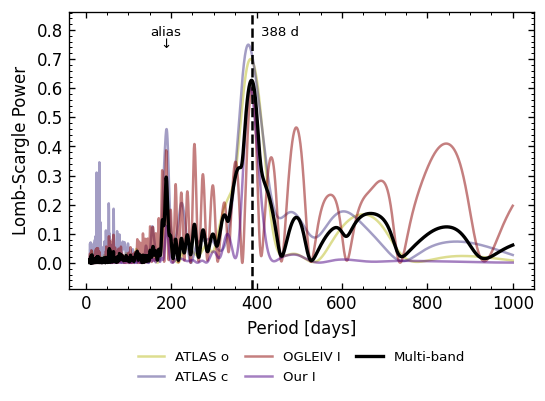}
 \includegraphics[width=0.469\textwidth]{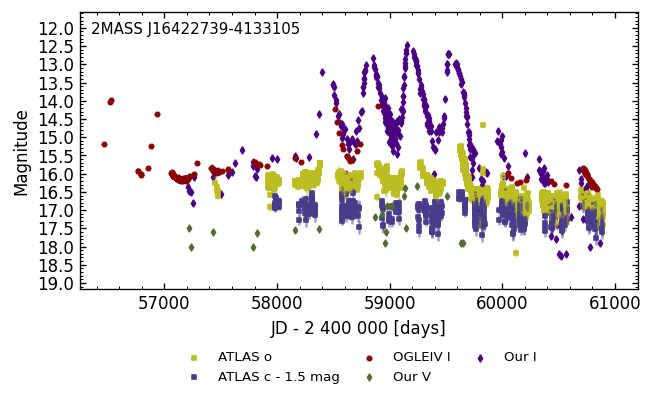}
 \includegraphics[width=0.40\textwidth]{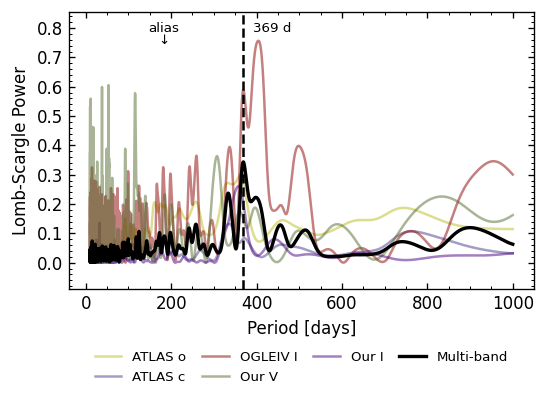}

 \caption{Light curves and Lomb–Scargle periodograms of the confirmed symbiotic stars from \citetalias{2014MNRAS.440.1410M}.}
 \label{fig:photometry8}
 \end{figure*}

     \begin{figure*}
\centering

 \includegraphics[width=0.469\textwidth]{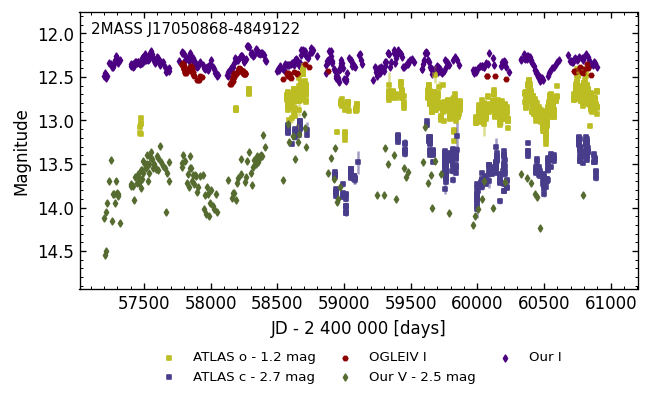}
 \includegraphics[width=0.40\textwidth]{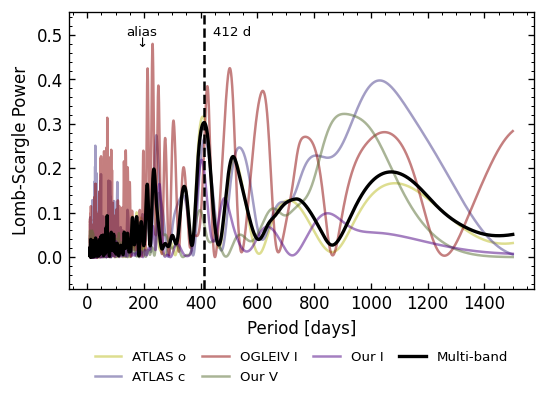}

  \includegraphics[width=0.469\textwidth]{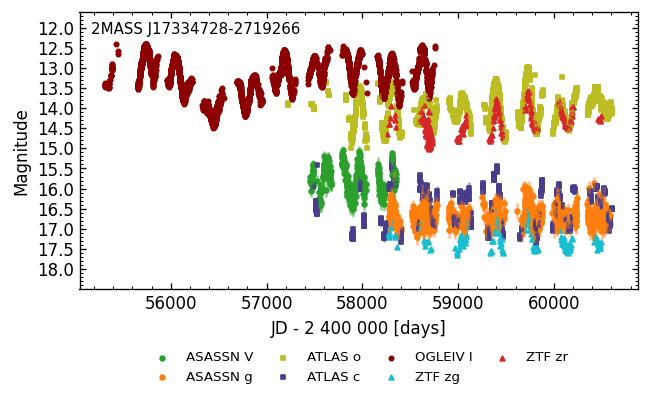}
 \includegraphics[width=0.40\textwidth]{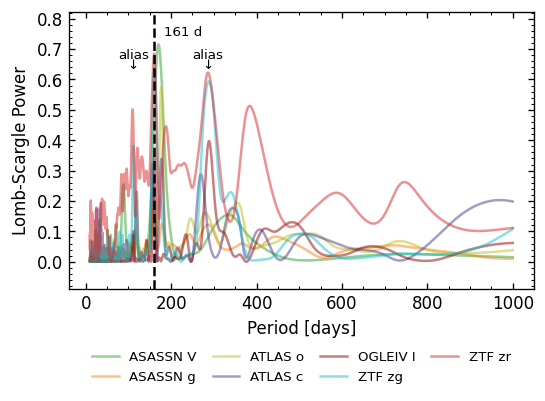}

 \includegraphics[width=0.469\textwidth]{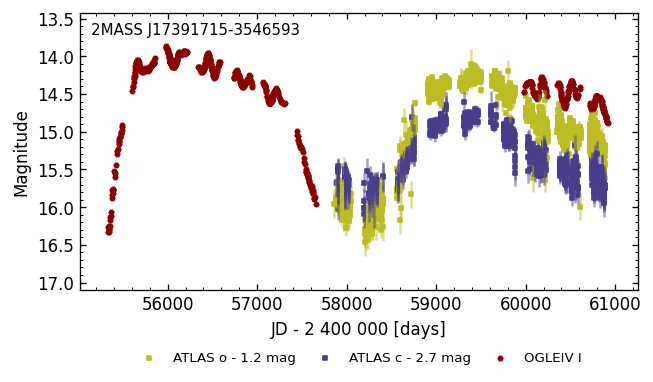}
 \includegraphics[width=0.40\textwidth]{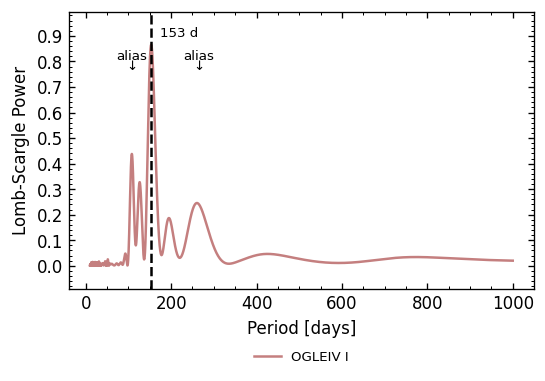}

  \includegraphics[width=0.469\textwidth]{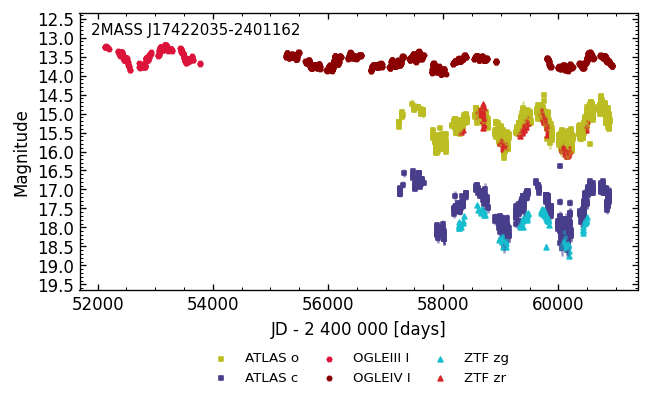}
 \includegraphics[width=0.40\textwidth]{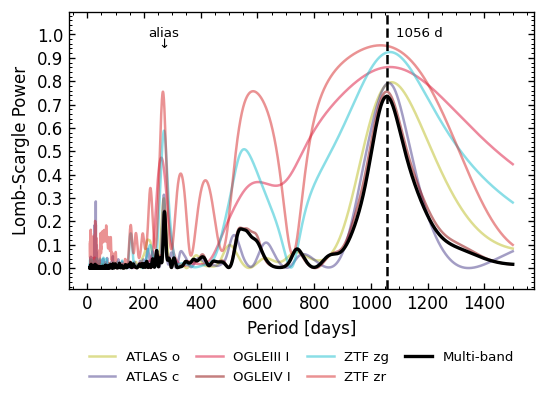}

 \contcaption{Light curves and Lomb–Scargle periodograms of the confirmed symbiotic stars from \citetalias{2014MNRAS.440.1410M}.}
 \label{fig:photometry9}
 \end{figure*}

      \begin{figure*}
\centering

 \includegraphics[width=0.469\textwidth]{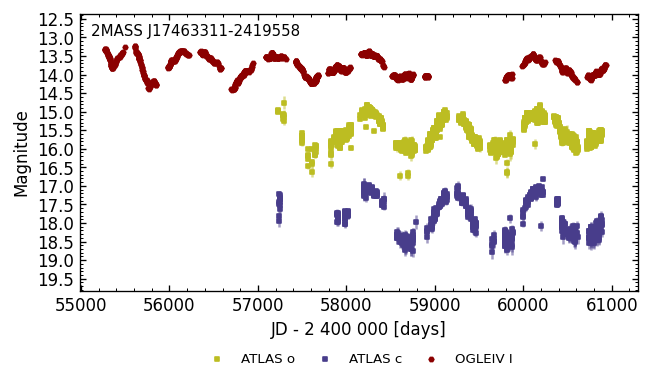}
 \includegraphics[width=0.40\textwidth]{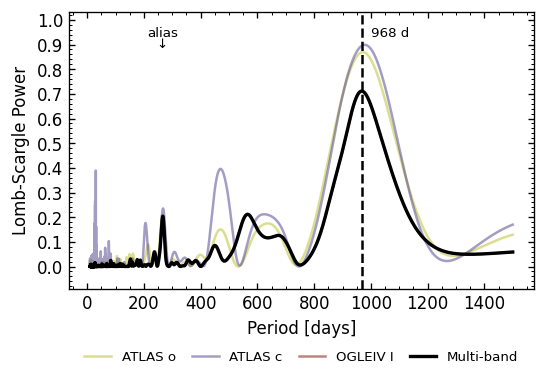}

  \includegraphics[width=0.469\textwidth]{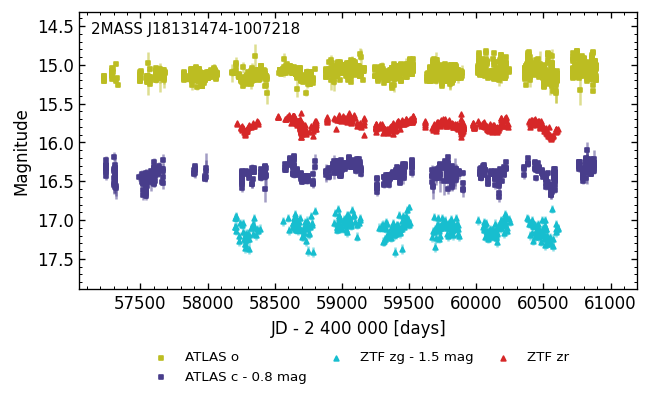}
 \includegraphics[width=0.40\textwidth]{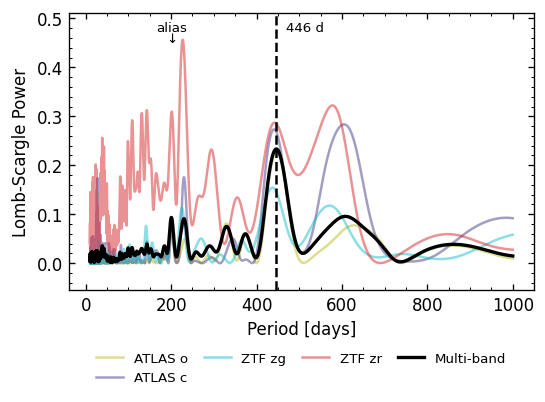}

 \includegraphics[width=0.469\textwidth]{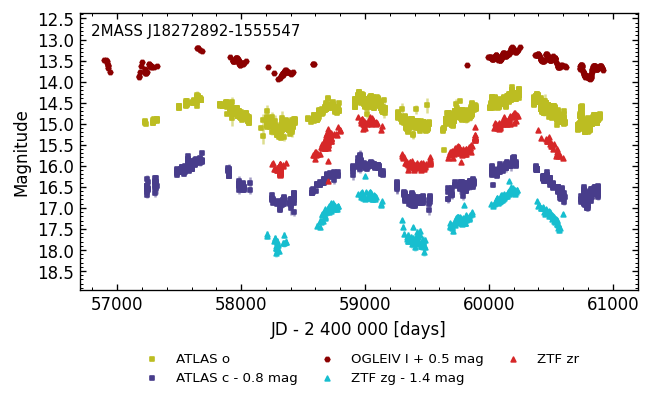}
 \includegraphics[width=0.40\textwidth]{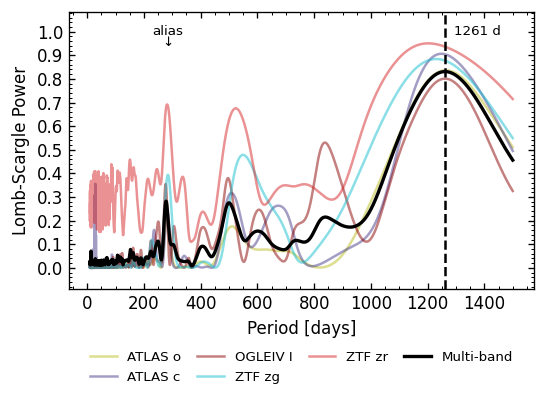}

  \includegraphics[width=0.469\textwidth]{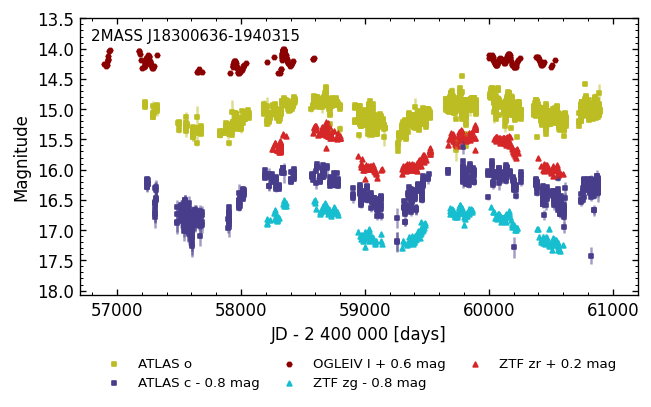}
 \includegraphics[width=0.40\textwidth]{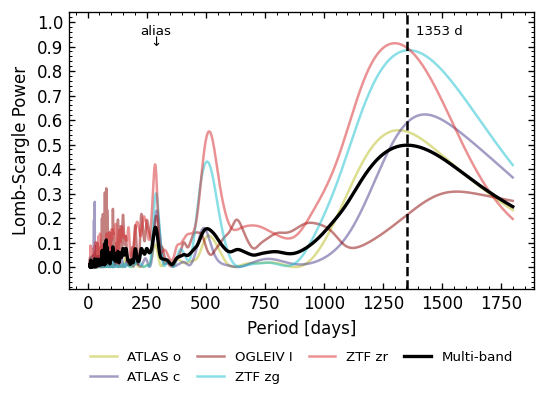}
 
 \contcaption{Light curves and Lomb–Scargle periodograms of the confirmed symbiotic stars from \citetalias{2014MNRAS.440.1410M}.}
 \label{fig:photometry11}
 \end{figure*}

       \begin{figure*}
\centering

 \includegraphics[width=0.469\textwidth]{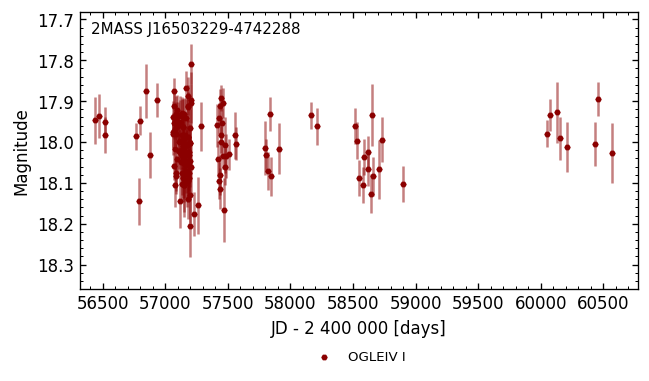}
 \includegraphics[width=0.40\textwidth]{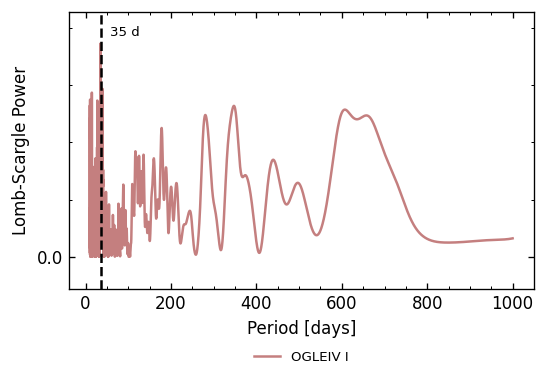}

  \includegraphics[width=0.469\textwidth]{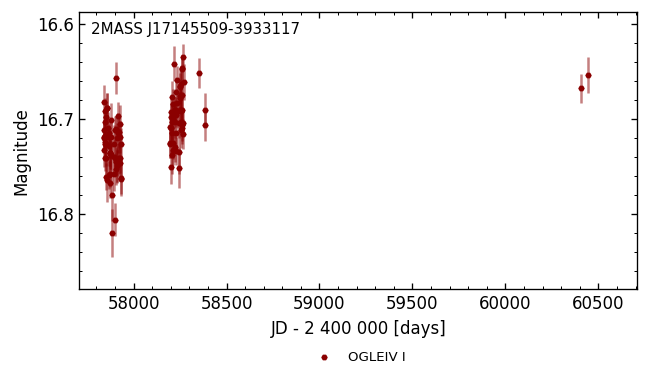}
 \includegraphics[width=0.40\textwidth]{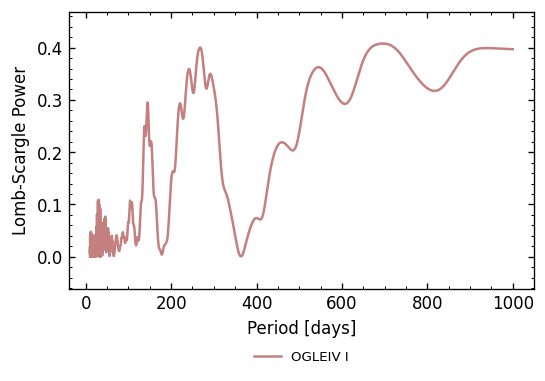}

\includegraphics[width=0.469\textwidth]{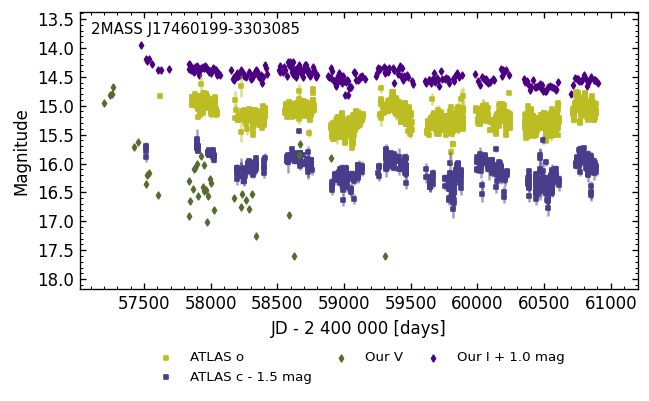}
 \includegraphics[width=0.40\textwidth]{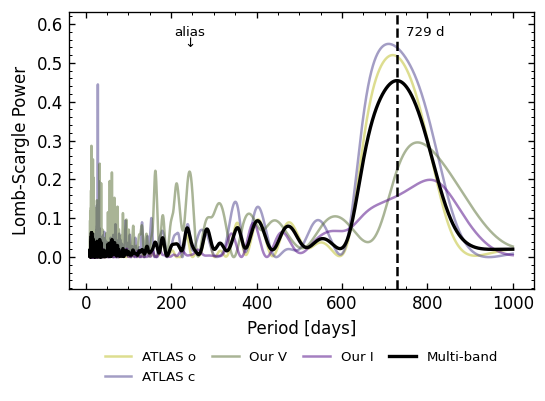}

 \caption{Light curves and Lomb–Scargle periodograms of the possible symbiotic stars from \citetalias{2014MNRAS.440.1410M}.}
 \label{fig:photometry10}
 \end{figure*}

\end{document}